\documentclass[pre,tighten]{revtex4}
\usepackage[cp850]{inputenc}
\usepackage{natbib}
\usepackage{amssymb,amsmath}
\usepackage{epsfig}
\usepackage{bm}
\usepackage{color}

\usepackage{graphicx}

\usepackage{color}

\newcommand\beq{\begin{equation}}
\newcommand\eeq{\end{equation}}
\newcommand\beqa{\begin{eqnarray}}
\newcommand\eeqa{\end{eqnarray}}
\newcommand{\dd}{\text{d}}

\newcommand{\al}{\alpha}

\begin{document}

\title{Kinetic theory of polydisperse granular mixtures: influence of
the partial temperatures on transport properties. A review}

\author{Mois\'es Garc\'{\i}a Chamorro\footnote[1]{Electronic address: moises@unex.es} and Rub\'en G\'omez Gonz\'alez\footnote[1]{Electronic address: ruben@unex.es}}
\affiliation{Departamento de F\'{\i}sica,
Universidad de Extremadura, E-06006 Badajoz, Spain}
\author{Vicente Garz\'{o}\footnote[2]{Electronic address: vicenteg@unex.es;
URL: http://www.unex.es/eweb/fisteor/vicente/}}
\affiliation{Departamento de F\'{\i}sica and Instituto de Computaci\'on Cient\'{\i}fica Avanzada (ICCAEx), Universidad de Extremadura, E-06006 Badajoz, Spain}



\begin{abstract}

It is well-recognized that granular media under rapid flow conditions can be modeled as a gas of hard spheres with inelastic collisions. At moderate densities, a fundamental basis for the determination of the granular hydrodynamics is provided by the Enskog kinetic equation conveniently adapted to account for inelastic collisions. A surprising result (compared to its molecular gas counterpart) for granular mixtures is the failure of the energy equipartition, even in homogeneous states. This means that the partial temperatures $T_i$ (measuring the mean kinetic energy of each species) are different to the (total) granular temperature $T$. The goal of this paper is to provide an overview on the effect of different partial temperatures on the transport properties of the mixture. Our analysis addresses first the impact of energy nonequipartition on transport which is only due to the inelastic character of collisions. This effect (which is absent for elastic collisions) is shown to be significant in important problems in granular mixtures such as thermal diffusion segregation. Then, an independent source of energy nonequipartition due to the existence of a divergence of the flow velocity is studied. This effect (which was already analyzed in several pioneering works on dense hard-sphere molecular mixtures) affects to the bulk viscosity coefficient. Analytical (approximate) results are compared against Monte Carlo and molecular dynamics simulations, showing the reliability of kinetic theory for describing granular flows.

\end{abstract}

\draft
\date{\today}
\maketitle

Keywords: Granular mixtures; Homogeneous cooling state; Enskog kinetic equation; Partial temperatures; DSMC method; Diffusion transport coefficients; Bulk viscosity coefficient.

\section{Introduction}
\label{sec1}

It is well-known that when granular matter is subjected to a violent and sustained excitation, the motion of grains resembles to the random motion of atoms or molecules in an ordinary or molecular gas. In this situation (referred usually to as rapid flow conditions), the energy injected to the system compensates for the energy dissipated by collisions and the effects of gravity. A system of activated collisional grains is referred to as a \emph{granular} gas; its study is the main objective of the present review.

Granular matter in nature is usually immersed in a fluid like water or air, so that a granular flow is a multiphase process. However, under some conditions (for instance, when the stress due to grains is larger than that exerted by the interstital fluid), the effect of the fluid phase on grains can be neglected. Here, we will address our attention to the study of the so-called \emph{dry} granular gases where the impact of the fluid phase on the dynamics of solid particles is not accounted for.

Since the grains which make up a granular material are of macroscopic size (their diameter is micrometers or larger), all the collisions among granular particles are \emph{inelastic}. This is one of the main differences to molecular gases. Due to this fact, the conventional methods of equilibrium statistical mechanics and thermodynamics fail. However, kinetic theory (which essentially addresses the dynamics of grains) is still an appropriate tool since it applies to elastic or inelastic collisions \cite{D00,D09a}. As we are mainly interested in assessing the effect of inelasticity of collisions on the dynamical properties of the granular particles, it is quite usual to consider a relatively simple (idealized) model which isolates the collisional dissipation effect from other relevant properties of granular matter. The most popular model for granular gases is a system of identical smooth hard spheres with a constant (positive) coefficient of normal restitution $\al \leqslant 1$. This quantity measures the ratio between the magnitude of the normal component of the relative velocity (oriented along the line separating the centers of the two spheres at contact) before and after a collision. The case $\al=1$ corresponds to perfectly elastic collisions while when $\al<1$ part of the kinetic energy of the relative motion is lost.

Within the context of the inelastic hard sphere model, the Boltzmann and Enskog kinetic equations have been conveniently extended  to account for the dissipative character of collisions \cite{GS95,BDS97,NEB98,D00,PL01,G03,BP04,RN08,D09a,G19}. While the Boltzmann equation applies to low-density gases, the Enskog equation holds for moderately dense gases. These kinetic equations have been employed in the last few years as the starting point to derive the corresponding \emph{granular} hydrodynamic equations. In particular, in the case of monocomponent granular gases and assuming the existence of a \emph{normal} (or hydrodynamic) solution for sufficiently long space and time scales, the Chapman--Enskog \cite{CC70} and Grad's moment \cite{G49} methods have been applied to solve the Boltzmann and Enskog kinetic equations to the Navier--Stokes order and obtain explicit expressions for the transport coefficients \cite{JS83,LSJCh84,JR85a,JR85b,BDKS98,GD99a,L05,G13}.

On the other hand, since a real granular system is usually characterized by some degree of polydispersity in density and size, flows of granular mixtures are prevalent in both nature and industry. For instance, natural systems that are highly polidisperse and propagate as rapid granular flows are pyroplastic density currents \cite{LBEDB20}, landslides and debris flows \cite{IRL97}, and rock avalanches \cite{H06}. Examples of industrial systems include mixing of pharmaceutical powders and poultry feedstock.

Needless to say, in the context of kinetic theory, the determination of the Navier--Stokes transport coefficients of a granular mixture is more intricate than that of a monocomponent granular gas since not only is the number of transport coefficients larger than for a single gas but also they depend on many parameters (masses and diameters, concentrations, and coefficients of restitution). Thus, due to this type of technical difficulties, many of the early attempts \cite{JM89,Z95,AW98,WA99} to obtain the transport coefficients of a granular mixture were carried out by assuming equipartition of energy: the partial temperatures $T_i$ of each species are equal to the (total) granular temperature $T$. A consequence of this assumption is that the Chapman--Enskog expansion was performed around Maxwellian distributions at the same temperature $T$ for each species. The use of this Maxwellian distribution as the reference state in the Chapman--Enskog method can be only considered as reliable for nearly elastic spheres where the energy equipartition still holds. Moreover, within this level of approximation, the expressions of the transport coefficients are the same as those obtained for molecular (elastic) mixtures \cite{FK72,CC70}; the inelasticity in collisions is only taken into account by the presence of a sink term in the energy balance equation.

However, many different works based on kinetic theory \cite{GD99b,MP99}, computer simulations \cite{MG02,BT02,DHGD02,PMP02,CH02,BT02b,KT03,WJM03,BRM05,AL05,SUKSS06,VLSG17,LVGS19,BSG20} and real experiments \cite{WP02,FM02} have clearly shown the failure of the energy equipartition in granular mixtures. This failure occurs even in homogeneous situations (in the so-called homogeneous cooling state) and is a consequence of both the inelasticity in collisions and the mechanical differences of the particles (e.g., masses, diameters). In fact, nonequipartition disappears when collisions between the different species of the mixture are elastic or when they are mechanically equivalent. Although the possibility of energy nonequipartition in granular mixtures was already noted by Jenkins and Mancini \cite{JM87}, to the best of our knowledge the impact of nonequipartition on transport properties in granular mixtures was computed for the first time by Huilin \emph{et al.} \cite{HWRLG00,HGM01}. However, these authors do not attempt to solve the kinetic equation and they assume local Maxwellian distribution functions for each species even in inhomogeneous states. Although this procedure can be employed to get the collisional transfer contributions to the fluxes, it predicts vanishing Navier--Stokes transport coefficients for dilute granular mixtures which is of course a wrong result. A more rigorous way of incorporating energy nonequipartition in the Chapman--Enskog solution has been published in the past few years \cite{GD02,GMD06,GM07,GDH07,GHD07}. The results have clearly shown that in general the effect of temperatures differences on the Navier--Stokes transport coefficients are important, specially for disparate masses or sizes and/or strong inelasticity.

As the Chapman--Enskog procedure states \cite{CC70}, since the partial temperatures are kinetic quantities, they must be also expanded in terms of gradients of the hydrodynamic fields. The partial temperatures are scalars so that, their first-order contributions $T_i^{(1)}$ must be proportional to the divergence of the flow velocity $\mathbf{U}$. Thus, a different way of inducing a breakdown of the energy equipartition in granular mixtures is by the presence of the gradient $\nabla \cdot \mathbf{U}$. This effect is not generic of granular mixtures since it was already found in the pioneering works of dense hard-sphere mixtures with elastic collisions \cite{LCK83,KS79a,KS79b}. The non-vanishing divergence of the mean flow velocity $\nabla \cdot \mathbf{U}$ causes that $T_i^{(1)}$ is involved in the evaluation of the bulk viscosity (proportionality coefficient between the collisional part of the pressure tensor and $\nabla\cdot \mathbf{U}$) as well as in the first-order contribution to the cooling rate $\zeta$ (which accounts for the rate of kinetic energy dissipated by collisions).

The aim of this paper is to offer a short review on the influence of the energy nonequipartition on transport properties in granular mixtures. Since we will consider moderate densities, the one-particle velocity distribution functions of each species will obey the set of coupled Enskog kinetic equations. The review is structured as follows. The set of Enskog coupled kinetic equations for a multicomponent granular mixture and its associated macroscopic balance equations are introduced in section \ref{sec2}. In particular, explicit forms for the collisional transfer contributions to the fluxes are given in terms of the one-particle velocity distribution function $f_i$ of each species. Section \ref{sec3} deals with the solution to the Enskog equation in the homogeneous cooling state; a homogeneous state where the granular temperature decreases in time due to inelastic cooling. As for monocomponent granular gases, a scaling solution is proposed in which the time dependence of the distributions $f_i$ occurs entirely through the temperature of the mixture $T$. The temperature ratios $T_i^{(0)}/T$ are determined by the condition of equal cooling rates $\zeta_i$. An approximate solution is obtained by truncating the expansion of the distributions $f_i$ in Sonine (or Laguerre) polynomials; the results show that $T_i^{(0)}/T\neq 1$ (energy nonequipartition). In section \ref{sec3}, the (approximate) theoretical results are compared against the results obtained from both the Direct Simulation Monte Carlo (DSMC) method and molecular dynamics (MD) simulations for conditions of practical interest. Comparison shows in general a good agreement between theory and simulations. The forms of the Navier--Stokes transport coefficients of the mixture in terms of the first-order distributions derived from the application of the Chapman--Enskog method around the local version of the homogeneous distributions obtained in section \ref{sec2} are displayed in section \ref{sec4}. Section \ref{sec5} addresses one of the main targets of the present paper: the study of the influence of the temperature ratios $T_i^{(0)}/T$ on the transport coefficients. To show more clearly the impact of nonequipartition on transport, we focus here on our attention to the diffusion transport coefficients of a dilute granular binary mixture. As expected, we find that the effect of energy nonequipartition on transport is in general quite significant. This means that, beyond nearly elastic systems, any reliable theory devoted to granular mixtures must include this nonequipartition effect. The influence of the first-order contributions $T_i^{(1)}$ to the partial temperatures on the bulk viscosity $\eta_\text{b}$ and the cooling rate $\zeta$ is widely analyzed in section \ref{sec6}. The contributions to $\eta_\text{b}$ coming from the coefficients $T_i^{(1)}$ were implicitly neglected in several previous works \cite{GDH07,GHD07,MGH12} on dense granular mixtures. Our present results indicate that the impact of $T_i^{(1)}$ on $\eta_\text{b}$ cannot be neglected for disparate masses and/or strong inelasticity. The paper is ended in section \ref{sec8} with a brief discussion of the results reported here.

Before ending this section, we want to remark that the present account is based on the authors' taste and perspective. In this sense, no attempt is made to include the extensive related work of many others in this field. The references given are selective and apologies are offered at the outset to the many other important contributions not recognized explicitly.

\section{Enskog kinetic equation for polydisperse dense granular mixtures}
\label{sec2}

\subsection{Enskog kinetic equation for inelastic hard spheres}

We consider a granular mixture of inelastic hard disks ($d=2$) or spheres ($d=3$) of masses $m_i$ and diameters $\sigma_i$ ($i=1,2,\ldots, s$). The subscript $i$ labels one of the $s$ mechanically different species or components and $d$ is the dimension of the system. For the sake of simplicity, we assume that the spheres are completely smooth; this means that inelasticity of collisions between particles of species $i$ and $j$ is only characterized by the constant (positive) coefficients of restitution $\al_{ij}\leqslant 1$. The coefficient $\al_{ij}$ measures the ratio between the magnitude of the \emph{normal} component (along the line separating the centers of the two spheres at contact) of the relative velocity after and before the collision $i$-$j$. 


For moderate densities, the one-particle velocity distribution function $f_i(\mathbf{r}, \mathbf{v}, t)$ of species $i$ verifies the set of $s$-coupled nonlinear integro-differential Enskog equations. In the absence of any external force, the Enskog kinetic equations are given by \cite{G19}
\beq
\label{2.1}
\frac{\partial f_i}{\partial t}+\mathbf{v}\cdot \nabla f_i=\sum_{j=1}^s\;J_{ij}[\mathbf{r},\mathbf{v}|f_i,f_j],\quad i=1,2,\ldots, s,
\eeq
where the Enskog collision operator $J_{ij}[\mathbf{r},\mathbf{v}|f_i,f_j]$ is
\beqa
\label{2.2}
J_{ij}\left[\mathbf{r}_1, \mathbf{v}_1|f_i,f_j\right]&=&\sigma_{ij}^{d-1}\int d\mathbf{v}_2\int d\widehat{\boldsymbol{\sigma}}\Theta\left(\widehat{\boldsymbol{\sigma}}\cdot\mathbf{g}_{12}\right)
\left(\widehat{\boldsymbol{\sigma}}\cdot\mathbf{g}_{12}\right)\Big[\alpha_{ij}^{-2}\chi_{ij}
(\mathbf{r}_1,\mathbf{r}_1-\boldsymbol{\sigma}_{ij})f_i(\mathbf{r}_1,\mathbf{v}_1'',t)\nonumber\\
& &
\times
f_j(\mathbf{r}_1-
\boldsymbol{\sigma}_{ij},\mathbf{v}_2'',t)
-\chi_{ij}(\mathbf{r}_1,\mathbf{r}_1+\boldsymbol{\sigma}_{ij})f_i(\mathbf{r}_1,\mathbf{v}_1,t)f_j(\mathbf{r}_1
+\boldsymbol{\sigma}_{ij},\mathbf{v}_2,t)\Big].
\eeqa
In Eq.\ \eqref{2.2}, $\boldsymbol{\sigma}_{ij}=\sigma_{ij} \widehat{\boldsymbol{\sigma}}$, $\sigma_{ij}=(\sigma_i+\sigma_j)/2$, $\widehat{\boldsymbol{\sigma}}$ is a unit vector directed along the line of centers from the sphere of species $i$ to that of species $j$ at contact, $\Theta$ is the Heaviside step function, and $\mathbf{g}_{12}=\mathbf{v}_1-\mathbf{v}_2$ is the relative velocity of the colliding pair. Moreover, $\chi_{ij}(\mathbf{r}_1,\mathbf{r}_1+\boldsymbol{\sigma}_{ij})$ is the equilibrium pair correlation function of two hard spheres, one of species $i$ and the other of species $j$ at contact, i.e., when the distance between their centers is $\sigma_{ij}$.


As in the case of elastic hard spheres, the interactions between inelastic hard spheres are modeled by instantaneous collisions where momentum is transferred along the line joining the centers of the two colliding spheres. 
The relationship between the pre-collisional velocities $(\mathbf{v}_1'',\mathbf{v}_2'')$ and the post-collisional velocities $(\mathbf{v}_1,\mathbf{v}_2)$ is
\beq
\label{2.3}
\mathbf{v}_1''=\mathbf{v}_1-\mu_{ji}\left(1+\alpha_{ij}^{-1}\right)
\left(\boldsymbol{\widehat{\sigma}}
\cdot\mathbf{g}_{12}\right)\boldsymbol{\widehat{\sigma}}, \quad \mathbf{v}_2''=\mathbf{v}_2+\mu_{ij}\left(1+\alpha_{ij}^{-1}\right)
\left(\boldsymbol{\widehat{\sigma}}
\cdot\mathbf{g}_{12}\right)\boldsymbol{\widehat{\sigma}},
\eeq
where $\mu_{ij}=m_i/(m_i+m_j)$. Equations \eqref{2.3} give the so-called inverse or \emph{restituting} collisions. Inversion of these collision rules provides the form of the so-called \emph{direct} collisions, namely, collisions where the pre-collisional velocities $(\mathbf{v}_1,\mathbf{v}_2)$ lead to the post-collisional velocities $(\mathbf{v}_1',\mathbf{v}_2')$ \cite{BP04}:
\beq
\label{2.4}
\mathbf{v}_1'=\mathbf{v}_1-\mu_{ji}\left(1+\alpha_{ij}\right)\left(\boldsymbol{\widehat{\sigma}}
\cdot\mathbf{g}_{12}\right)\boldsymbol{\widehat{\sigma}}, \quad \mathbf{v}_2'=\mathbf{v}_2+\mu_{ij}\left(1+\alpha_{ij}\right)\left(\boldsymbol{\widehat{\sigma}}
\cdot\mathbf{g}_{12}\right)\boldsymbol{\widehat{\sigma}}.
\eeq
From Eqs.\ \eqref{2.3} and \eqref{2.4}, one gets the relations
\beq
\label{2.4.1}
\left(\boldsymbol{\widehat{\sigma}}\cdot\mathbf{g}_{12}''\right)=-\al_{ij}^{-1} \left(\boldsymbol{\widehat{\sigma}}
\cdot\mathbf{g}_{12}\right), \quad \left(\boldsymbol{\widehat{\sigma}}\cdot\mathbf{g}_{12}'\right)=-\al_{ij} \left(\boldsymbol{\widehat{\sigma}}
\cdot\mathbf{g}_{12}\right),
\eeq
where $\mathbf{g}_{12}''=\mathbf{v}_1''-\mathbf{v}_2''$ and $\mathbf{g}_{12}'=\mathbf{v}_1'-\mathbf{v}_2'$. For inelastic collisions, it is quite apparent from Eq.\ \eqref{2.4.1} that the magnitude of the normal component of the pre-collisional relative velocity is larger than its post-collisional counterpart. In addition,  comparison between Eqs.\ \eqref{2.3} and \eqref{2.4} shows that, except for molecular mixtures (elastic collisions), the direct and inverse collisions are not equivalent. This is essentially due to the lack of time reversal symmetry for inelastic collisions.

The change in kinetic energy of the colliding pair in a binary collision can be easily obtained from Eq.\ \eqref{2.4}:
\beqa
\label{2.4.2}
\Delta E_{ij}\equiv E_{ij}'-E_{ij}&=&\frac{m_i}{2}v_1^{'2}+\frac{m_j}{2}v_2^{'2}-\left(\frac{m_i}{2}v_1^{2}+\frac{m_j}{2}v_2^{2}\right)\nonumber\\
&=&-\frac{m_{ij}}{2}\left(\boldsymbol{\widehat{\sigma}}\cdot\mathbf{g}_{12}\right)^2\left(1-\al_{ij}^2\right),
\eeqa
where $m_{ij}=m_im_j/(m_i+m_j)$ is the reduced mass. When $\al_{ij}=1$ (elastic collisions), Eq.\ \eqref{2.4.2} leads to $\Delta E_{ij}=0$, as expected for molecular mixtures. When $\al_{ij}<1$, $\Delta E_{ij}< 0$ so that, part of the kinetic energy is lost in a binary collision between a particle of species $i$ and a particle of species $j$.

\subsection{Macroscopic balance equations}

The knowledge of the velocity distribution functions $f_i$ allows us to obtain the hydrodynamic fields of the multicomponent mixture. The quantities of interest in a macroscopic description of the granular mixture are the local number density $n_i$ of species $i$, the local mean flow velocity of the mixture $\mathbf{U}$, and the granular temperature $T$. In terms of the distributions $f_i$, they are defined, respectively, as
\beq
\label{2.5}
n_i(\mathbf{r};t)=\int d \mathbf{v}\; f_i(\mathbf{r}, \mathbf{v}; t),
\eeq
\beq
\label{2.6}
\mathbf{U}(\mathbf{r};t)=\frac{1}{\rho(\mathbf{r};t)}\sum_{i=1}^s \int d \mathbf{v}\; m_i \mathbf{v} f_i(\mathbf{r}, \mathbf{v}; t),
\eeq
\beq
\label{2.7}
T(\mathbf{r};t)=\frac{1}{n(\mathbf{r};t)}\sum_{i=1}^s\int d\mathbf{v}\frac{m_{i}}{d}V^{2}f_{i}(\mathbf{r}, \mathbf{v}; t).
\eeq
In Eqs.\ \eqref{2.5}--\eqref{2.7}, $\rho=\sum_i m_i n_i$ is the total mass density, $\rho_i=m_i n_i$ is the mass density of the species $i$,  $n=\sum_i n_i$ is the total number density, and $\mathbf{V}=\mathbf{v}-\mathbf{U}$ is the peculiar velocity. For the subsequent discussion, at a kinetic level, it is convenient to introduce the partial kinetic temperatures $T_i$ for each species. The temperature $T_i$ provides a measure of the mean kinetic energy of the species $i$. The partial temperatures are defined as
\begin{equation}
\label{2.8}
T_i(\mathbf{r};t)=\frac{m_{i}}{d n_i(\mathbf{r};t)}\int\; d\mathbf{v}\;V^{2}f_{i}(\mathbf{r}, \mathbf{v}; t).
\end{equation}
From Eqs.\ \eqref{2.7} and \eqref{2.8}, the granular temperature $T$ of the mixture can be also written in terms of the partial temperatures $T_i$ as
\beq
\label{2.9}
T(\mathbf{r};t)=\sum_{i=1}^s\, x_i(\mathbf{r};t) T_i(\mathbf{r};t),
\eeq
where $x_i=n_i/n$ is the concentration or mole fraction of species $i$. Thus, due to the constraint \eqref{2.9}, there are $s-1$ independent partial temperatures in a mixture constituted by $s$ components.

As occurs for molecular mixtures \cite{FK72}, the fields $n_i$, $\mathbf{U}$, and $T$ are expected to be the \emph{slow} variables that dominate the dynamics of the mixture for sufficiently long times through the set of hydrodynamic equations. For elastic collisions, the above fields are the densities of global conserved quantities and so, they persist at long times (in comparison with the mean free time) where the complex microscopic dynamics becomes negligible \cite{RL77,C88}. In the case of granular fluids, the energy is not conserved in collisions and the rate of energy dissipated by collisions is characterized (as we will see below) by a cooling rate. However, as confirmed by MD simulations (see, for instance, Ref.\ \cite{DHGD02}), the cooling rate may be slow compared to the transient dynamics so that, the kinetic energy (or granular temperature) can be still considered as a slow variable.

The balance equations for $n_i$, $\mathbf{U}$, and $T$ can be obtained by multiplying the set of Enskog equations \eqref{2.1} by $1$, $m_i\mathbf{v}$, and $\frac{m_i}{2}V^2$ and summing over all the species in the momentum and energy equations. The result is
\beq
\label{2.10}
D_t n_i+n_i\nabla\cdot \mathbf{U}+\frac{\nabla\cdot\mathbf{j}_i}{m_i}=0,
\eeq
\beq
\label{2.11}
\rho D_t\mathbf{U}+\nabla\cdot\mathsf{P}^\text{k}=\sum_{i=1}^s\sum_{j=1}^s\; \int d\mathbf{v}\; m_i \mathbf{v} J_{ij}[\mathbf{r}|f_i,f_j],
\eeq
\beq
\label{2.12}
D_tT-\frac{T}{n}\sum_{i=1}^s\frac{\nabla\cdot\mathbf{j}_i}{m_i}+\frac{2}{dn}
\left(\nabla\cdot\mathbf{q}^\text{k}+\mathsf{P}^\text{k}:\nabla\mathbf{U}\right)=\sum_{i=1}^s\sum_{j=1}^s\; \int d\mathbf{v}\; \frac{m_i}{2}V^2 J_{ij}[\mathbf{r}|f_i,f_j],
\eeq
where $D_t=\partial_t+\mathbf{U}\cdot\nabla$ is the  material derivative. In the above equations,
\beq
\label{2.13}
\mathbf{j}_i=m_i\int\;d\mathbf{v}\; \mathbf{V}f_i(\mathbf{v})
\eeq
is the mass flux for species $i$ relative to the local flow $\mathbf{U}$ and the kinetic contributions $\mathsf{P}^\text{k}$ and $\mathbf{q}^\text{k}$ to the pressure tensor and heat flux are given, respectively, by
\beq
\label{2.14}
\mathsf{P}^\text{k}=\sum_{i=1}^s\int d\mathbf{v}\; m_i\mathbf{V}\mathbf{V}f_i(\mathbf{v}),
\eeq
\beq
\label{2.15}
\mathbf{q}^\text{k}=\sum_{i=1}^s\int d\mathbf{v}\; \frac{m_i}{2}V^2\mathbf{V}f_i(\mathbf{v}).
\eeq
A consequence of the definition \eqref{2.13} of the fluxes $\mathbf{j}_i$ is that only $s-1$ mass fluxes are independent since they have the constraint
\beq
\label{2.15.1}
\sum_{i=1}^s\; \mathbf{j}_i=\mathbf{0}.
\eeq
Needless to say, to end the derivation of the balance hydrodynamic equations one has to compute the right-hand side of Eqs.\ \eqref{2.11} and \eqref{2.12}. These terms can be obtained by employing an important property of the integrals involving the Enskog collision operator $J_{ij}[\mathbf{r},\mathbf{v}|f_i,f_j]$ \cite{G19,GDH07}:
\beqa
\label{2.16}
I_{\psi_i}&\equiv & \sum_{i=1}^s\sum_{j=1}^s\int\; d \mathbf{v}_1\; \psi_i(\mathbf{v}_1) J_{ij}[\mathbf{r}_1,\mathbf{v}_1|f_i,f_j]\nonumber\\
&=&\frac{1}{2}\sum_{i=1}^s\sum_{j=1}^s\sigma_{ij}^{d-1}\int d\mathbf{v}_1\int\ d{\bf v}_{2}\int d\widehat{\boldsymbol{\sigma}}\,
\Theta (\widehat{{\boldsymbol {\sigma }}}\cdot \mathbf{g}_{12})
(\widehat{\boldsymbol {\sigma }}\cdot \mathbf{g}_{12})\nonumber\\
& & \times \Bigg\{\Big[\psi_i(\mathbf{v}_1')
+\psi_j(\mathbf{v}_2')-\psi_i(\mathbf{v}_1)-\psi_j(\mathbf{v}_2)\Big]
f_{ij}\left(\mathbf{r}_1,\mathbf{v}_1,\mathbf{r}_1+\boldsymbol{\sigma}_{ij},\mathbf{v}_2;t\right)+\frac{\partial}{\partial \mathbf{r}_1}\cdot \boldsymbol{\sigma}_{ij}\nonumber\\
& & \times \Big[\psi_i(\mathbf{v}_1')-\psi_i(\mathbf{v}_1)\Big]\int_0^1\; dx\; f_{ij}\Big(\mathbf{r}_1-x\boldsymbol{\sigma}_{ij},\mathbf{v}_1,\mathbf{r}_1+(1-x)\boldsymbol{\sigma}_{ij},\mathbf{v}_2;t\Big)\Bigg\},
\nonumber\\
\eeqa
where $\psi_i(\mathbf{v}_1)$ is an arbitrary function of $\mathbf{v}_1$, $\mathbf{v}_1'$ is defined by Eq.\ \eqref{2.4} and
\beq
\label{2.17}
f_{ij}\left(\mathbf{r}_1,\mathbf{v}_1,\mathbf{r}_2,\mathbf{v}_2;t\right)\equiv
\chi_{ij}(\mathbf{r}_1,\mathbf{r}_2) f_i(\mathbf{r}_1,\mathbf{v}_1,t)
f_j(\mathbf{r}_2,\mathbf{v}_2,t).
\eeq
The first term on the right hand side of Eq.\ \eqref{2.16} represents a collisional effect due to scattering with a change in velocities. This term vanishes for elastic collisions. The second term on the right hand side of Eq.\ \eqref{2.16} provides a pure collisional effect due to the spatial difference of the colliding pair. This term vanishes for low-density mixtures.

In the case $\psi=m_i\mathbf{v}$, the first term in the integrand \eqref{2.16} disappears since the momentum is conserved in all pair collisions, i.e., $m_i \mathbf{v}_1+m_j \mathbf{v}_2=m_i \mathbf{v}_1'+m_j \mathbf{v}_2'$. The second term in the integrand yields the result
\beq
\label{2.18}
I_{p}\equiv \sum_{i=1}^s\sum_{j=1}^s\int\; d \mathbf{v}_1\; m_i \mathbf{v}_1 J_{ij}[\mathbf{r}_1,\mathbf{v}_1|f_i,f_j]=-\nabla \cdot \mathsf{P}^\text{c},
\eeq
where the collision transfer contribution to the pressure tensor $\mathsf{P}^\text{c}$ is \cite{GDH07}
\beqa
\label{2.19}
\mathsf{P}^\text{c}&=&\sum_{i=1}^s\sum_{j=1}^s\sigma_{ij}^d m_{ij}\frac{1+\alpha_{ij}}{2}\int d\mathbf{v}_1\int d\mathbf{v}_2
\int d\widehat{\boldsymbol{\sigma}}\Theta\left(\widehat{\boldsymbol{\sigma}}\cdot\mathbf{g}_{12}\right)\left(\widehat{\boldsymbol{\sigma}}
\cdot\mathbf{g}_{12}\right)^2\widehat{\boldsymbol{\sigma}}\widehat{\boldsymbol{\sigma}}\nonumber\\
& & \times \int_{0}^{1}dx f_{ij}\Big(\mathbf{r}-x\boldsymbol{\sigma}_{ij},\mathbf{r}+(1-x)\boldsymbol{\sigma}_{ij},\mathbf{v}_1,\mathbf{v}_2,t\Big).
\eeqa
In the case $\psi=\frac{1}{2}m_i V^2$, the first term on the right hand side of Eq.\ \eqref{2.16} does not vanish since the kinetic energy is not conserved in collisions. As before, the second term in the integrand gives the collisional transfer contribution to the heat flux $\mathbf{q}^\text{c}$. The result is \cite{GDH07}
\beq
\label{2.20}
I_{e}\equiv \sum_{i=1}^s\sum_{j=1}^s\int\; d \mathbf{v}_1\; \frac{1}{2}m_i V_1^2 J_{ij}[\mathbf{r}_1,\mathbf{v}_1|f_i,f_j]=-\nabla\cdot \mathbf{q}^\text{c} -\mathsf{P}^\text{c}:\nabla \mathbf{U}-\frac{d}{2}n T \zeta,
\eeq
where
\beqa
\label{2.21}
\mathbf{q}^\text{c}&=&\sum_{i=1}^s\sum_{j=1}^s\sigma_{ij}^dm_{ij}\frac{1+\alpha_{ij}}{8}\int d\mathbf{v}_1\int d\mathbf{v}_2\int d\widehat{\boldsymbol{\sigma}}\Theta\left(\widehat{\boldsymbol{\sigma}}\cdot\mathbf{g}_{12}\right)\left(\widehat{\boldsymbol{\sigma}}
\cdot\mathbf{g}_{12}\right)^2\widehat{\boldsymbol{\sigma}}\Big[4\left(\widehat{\boldsymbol{\sigma}}
\cdot\mathbf{G}_{ij}\right) \nonumber\\
& & +
\left(\mu_{ji}-\mu_{ij}\right)\left(1-\alpha_{ij}\right)\left(\boldsymbol{\hat{\sigma}}\cdot\mathbf{g}_{12}\right)
\Big]\int_{0}^{1}d x\; f_{ij}\Big(\mathbf{r}-x\boldsymbol{\sigma}_{ij},\mathbf{r}+(1-x)
\boldsymbol{\sigma}_{ij},\mathbf{v}_1,\mathbf{v}_2;t\Big),
\nonumber\\
\eeqa
and the (total) cooling rate $\zeta$ due to inelastic collisions among all species is given by
\beqa
\label{2.22}
\zeta&=&\frac{1}{2dnT}\sum_{i=1}^s\sum_{j=1}^s\sigma_{ij}^{d-1}m_{ij}\left(1-\alpha_{ij}^2\right)\int d\mathbf{v}_1\int d\mathbf{v}_2\int d\widehat{\boldsymbol{\sigma}}\Theta\left(\widehat{\boldsymbol{\sigma}}\cdot\mathbf{g}_{12}\right)\left(\widehat{\boldsymbol{\sigma}}
\cdot\mathbf{g}_{12}\right)^3\nonumber\\
& & \times
f_{ij}\left(\mathbf{r},\mathbf{r}+\boldsymbol{\sigma}_{ij},\mathbf{v}_1,\mathbf{v}_2;t\right).
\eeqa

The balance hydrodynamic equations for the densities of momentum and energy can be finally written when Eqs.\ \eqref{2.18}--\eqref{2.22} are substituted into the right hand sides of Eqs.\ \eqref{2.11} and \eqref{2.12}. These balance equations can be written as
\beq
\label{2.23}
D_t\mathbf{U}+\rho^{-1}\nabla\cdot\mathsf{P}=\mathbf{0},
\eeq
\beq
\label{2.24}
D_tT-\frac{T}{n}\sum_{i=1}^s\frac{\nabla\cdot\mathbf{j}_i}{m_i}+\frac{2}{dn}
\left(\nabla\cdot\mathbf{q}+\mathsf{P}:\nabla\mathbf{U}\right)=-\zeta T,
\eeq
where the pressure tensor $\mathsf{P}(\mathbf{r},t)$ and the heat flux $\mathbf{q}(\mathbf{r},t)$ have both kinetic and collisional transfer contributions, i.e.,
\beq
\label{4.13}
\mathsf{P}=\mathsf{P}^\text{k}+\mathsf{P}^\text{c}, \quad \mathbf{q}=\mathbf{q}^\text{k}+\mathbf{q}^\text{c}.
\eeq

Equations \eqref{2.10}, \eqref{2.23} and \eqref{2.24} are the balance equations for the hydrodynamic fields $n_i$, $\mathbf{U}$, and $T$, respectively, of a polydisperse granular mixture at moderate densities. This set of equations do not constitute a closed set of equations unless one expresses the fluxes and the cooling rate in terms of the above hydrodynamic fields and their spatial gradients. For small gradients, the corresponding constitutive equations for the fluxes and the cooling rate can be obtained by solving the set of Enskog kinetic equations \eqref{2.1} with the extension of the conventional Chapman--Enskog method \cite{CC70} to dissipative dynamics.

Before closing this section, it is instructive to consider the case of \emph{dilute} polydisperse granular mixtures. The corresponding balance equations can be obtained from Eqs.\ \eqref{2.10}, \eqref{2.23} and \eqref{2.24} by taking $\chi_{ij}\to 1$ and neglecting the different centers [$(\mathbf{r},\mathbf{r}\pm \boldsymbol{\sigma}_{ij})$] of the colliding pair since the effective diameter $\sigma_{ij}$ is much smaller than that of the mean free path of this collision. This implies that the collision transfer contributions to the fluxes are much smaller than their corresponding kinetic counterparts ($\mathsf{P} \to \mathsf{P}^\text{k}$ and $\mathbf{q} \to \mathbf{q}^\text{k}$) and the cooling rate $\zeta$ is simply
given by
\beqa
\label{4.14}
\zeta(\mathbf{r};t)&=&\frac{1}{2dnT}\sum_{i=1}^s\sum_{j=1}^s\sigma_{ij}^{d-1}m_{ij}\left(1-\alpha_{ij}^2\right)\int d\mathbf{v}_1\int d\mathbf{v}_2\int d\widehat{\boldsymbol{\sigma}}\Theta\left(\widehat{\boldsymbol{\sigma}}\cdot\mathbf{g}_{12}\right)
\left(\widehat{\boldsymbol{\sigma}}
\cdot\mathbf{g}_{12}\right)^3 \nonumber\\
& & \times f_i(\mathbf{r},\mathbf{v}_1;t)f_j(\mathbf{r},\mathbf{v}_2;t).
\eeqa

\section{Homogeneous cooling state. Partial temperatures}
\label{sec3}

We consider a spatially homogeneous state of an isolated polydisperse granular mixture. In contrast to molecular mixtures, there is no longer an evolution toward the local Maxwellian distributions since those distributions are not a solution to the set of homogeneous (inelastic) Enskog equations. Instead, as we will show below, there is an special solution which is achieved after a few collision times by considering homogeneous initial conditions: the so-called homogeneous cooling state (HCS).

For spatially homogeneous isotropic states, the set of Enskog equations for the distributions $f_i(v;t)$ reads
\beq
\label{3.1}
\frac{\partial}{\partial t}f_i(v;t)=\sum_{j=1}^s \chi_{ij} J_{ij}^\text{B}[\mathbf{v}|f_i,f_j],
\eeq
where the Boltzmann collision operator $J_{ij}^\text{B}$ is
\beq
\label{3.2}
J_{ij}^\text{B}[\mathbf{v}|f_i,f_j]=\sigma_{ij}^{d-1}\int d\mathbf{v}_2\int d\widehat{\boldsymbol{\sigma}}\Theta\left(\widehat{\boldsymbol{\sigma}}\cdot\mathbf{g}_{12}\right)\left(\widehat{\boldsymbol{\sigma}}
\cdot\mathbf{g}_{12}\right)\left[\al_{ij}^{-2}f_i(v_1'';t)f_j(v_2'';t)- f_i(v_1;t)f_j(v_2;t)\right].
\eeq
Upon writing Eqs.\ \eqref{3.1} and \eqref{3.2} we have taken into account that the dependence of the distributions $f_i$ on velocity $\mathbf{v}$ is only through its magnitude $v$. In Eq.\ \eqref{3.1}, note that $\chi_{ij}$ refers to the pair correlation function for particles of species $i$ and $j$ when they are separated a distance $\sigma_{ij}$.

For homogeneous states, the balance equations  \eqref{2.10} and \eqref{2.23} trivially hold. On the other hand, the balance equation of the granular temperature \eqref{2.24} yields
\beq
\label{3.3}
\frac{\partial T}{\partial t}=-T \zeta,
\eeq
where the cooling rate $\zeta$ is defined in Eq.\ \eqref{2.22} by making the replacement $f_{ij}(\mathbf{r}_1, \mathbf{v}_1, \mathbf{r}_2, \mathbf{v}_2; t)\to \chi_{ij}f_i(v_1;t)f_j(v_2;t)$. On the other hand, for homogeneous states, the integration in $\widehat{\boldsymbol{\sigma}}$ can be easily performed and $\zeta$ can be more explicitly written as
\beq
\label{3.4}
\zeta(t)=\frac{\pi^{(d-1)/2}}{2d\Gamma\left(\frac{d+3}{2}\right)}
\frac{1}{nT}\sum_{i=1}^s\sum_{j=1}^s\sigma_{ij}^{d-1}\chi_{ij} m_{ij}\left(1-\alpha_{ij}^2\right)\int d\mathbf{v}_1\int d\mathbf{v}_2 {g}_{12}^3 f_i({v}_1;t)f_j({v}_2;t).
\eeq
Moreover, for symmetry reasons, the mass and heat fluxes vanish and the pressure tensor $P_{k\ell}=p\delta_{k\ell}$, where the hydrostatic pressure $p$ is \cite{GDH07}
\beq
\label{3.5}
p=n T \Big[1+\frac{\pi^{d/2}}{d\Gamma\left(\frac{d}{2}\right)}\sum_{i=1}^s\sum_{j=1}^s\; \mu_{ji}\; n\; \sigma_{ij}^d \chi_{ij} x_i x_j (1+\al_{ij})\gamma_i\Big],
\eeq
where $\gamma_i=T_i^{(0)}/T$. Here, $T_i^{(0)}$ denotes the partial temperature of species $i$ in the homogeneous state (absence of spatial gradients).

To analyze the rate of change of the partial temperatures $T_i^{(0)}$, it is convenient to introduce the ``partial cooling rates'' $\zeta_i$. The definition of these quantities can be obtained by multiplying both sides of the Enskog equation \eqref{3.1} by $\frac{m_i}{2}v^2$ and integrating over velocity. The result is
\beq
\label{3.6}
\frac{\partial T_i^{(0)}}{\partial t}=-\zeta_i T_i^{(0)},
\eeq
where
\beq
\label{3.7}
\zeta_{i}(t)=\sum_{j=1}^s\zeta_{ij}, \quad \zeta_{ij}=-\frac{\chi_{ij}}{d n_i T_i^{(0)}}\int d\mathbf{v} m_i v^2 J_{ij}^\text{B}[f_i,f_j].
\eeq
From Eqs.\ \eqref{2.9}, \eqref{3.3}, and \eqref{3.6}, one can express the total cooling rate $\zeta$ in terms of the partial cooling rates $\zeta_i$ as
\beq
\label{3.8}
\zeta(t)=\sum_{i=1}^s\; x_i \gamma_i(t) \zeta_i(t).
\eeq
The time evolution of the temperature ratios $\gamma_i(t)$ can be easily derived from Eqs.\ \eqref{3.3} and \eqref{3.6} as
\beq
\label{3.9}
\frac{\partial \gamma_i}{\partial t}=\gamma_i\left(\zeta-\zeta_i\right).
\eeq

The term $\zeta_{ii}$ gives the contribution to the partial cooling rate $\zeta_i$ coming from the rate of energy loss from collisions between particles of the same species $i$. This term vanishes for elastic collisions but is different from zero when $\al_{ii}<1$. The remaining contributions $\zeta_{ij}$ ($i\neq j$) to $\zeta$ represent the transfer of energy between a particle of species $i$ and particles of species $j$. In general, the term $\zeta_{ij}\neq 0$  ($i\neq j$) for both elastic and inelastic collisions. However, in the special state where the distribution functions $f_i$ are Maxwellian distributions at the same temperature ($T_i^{(0)}=T$ for any species $i$), then $\zeta_{ij}=0$  ($i\neq j$) for elastic collisions. This is a consequence of the detailed balance for which the energy transfer between different species is balanced by the energy conservation for this state \cite{GD99b}.

The corresponding detailed balance state for inelastic collisions is the HCS. In this state, since the partial $\zeta_i$ and total $\zeta$ cooling rates never vanish, the partial $T_i^{(0)}$ and total $T$ temperatures are always time dependent. As for monocomponent granular gases \cite{BP04,NE98}, whatever the initial uniform state considered is, we expect that the Enskog equation \eqref{3.1} tends toward the HCS solution where all the time dependence of the distributions $f_i(v;t)$ only occurs through the (total) temperature $T(t)$. In this sense, the HCS solution qualifies as a \emph{normal} or hydrodynamic solution since the granular temperature $T(t)$ is in fact the relevant temperature at a hydrodynamic level. Thus, it follows from dimensional analysis that the distributions $f_i(v;t)$ have the form \cite{GD99b}
\beq
\label{3.10}
f_i(v;t)=n_i v_\text{th}^{-d}(t)\varphi_i \left(\frac{v}{v_\text{th}(t)}\right),
\eeq
where $v_\text{th}(t)=\sqrt{2T(t)/\overline{m}}$ is a thermal velocity defined in terms of the global temperature $T(t)$ of the mixture, $\overline{m}=(\sum_i\; m_i)/s$, and $\varphi_i$ is a reduced distribution function whose dependence on the (global) granular temperature $T(t)$ is through the dimensionless velocity $v/v_\text{th}(t)$.

Since the time dependence of the HCS solution \eqref{3.10} for $f_i$ only occurs through the (global) temperature $T(t)$, then the temperature ratios $\gamma_i$ must be \emph{independent} of time. This means that
all partial temperatures $T_i^{(0)}(t)$ are proportional to the (global) granular temperature $T(t)$ [$T_i^{(0)}(t)=\gamma_i T(t)$] and so,  the temperatures of the species do not provide any new dynamical degree of freedom at the hydrodynamic stage. However, they still characterize the shape of the velocity
distribution functions of each species and affect the quantitative averages (mass, momentum, and heat fluxes) calculated with these distributions.

As the temperature ratios do not depend on time, one possibility would be that $T_1^{(0)}=\ldots=T_s^{(0)}=T$, as happens in the case of molecular mixtures (elastic collisions). However, the ratios $T_i^{(0)}/T$ ($i=1,\ldots,s$) must be determined by solving the set of Enskog equations \eqref{3.1}. As we will show latter, the above ratios are in general different from 1 and exhibit a complex dependence on the parameter space of the mixture.

Since $\gamma_i\equiv \text{const}$, according to Eq.\ \eqref{3.9}, the partial cooling rates $\zeta_i$ must be equal in the HCS:
\beq
\label{3.11}
\zeta_1(t)=\zeta_2(t)=\cdots =\zeta_s(t)=\zeta(t).
\eeq
In addition, the right hand side of Eq.\ \eqref{2.1} can be more explicitly written when one takes into account Eq.\ \eqref{3.10}:
\beq
\label{3.12}
\frac{\partial f_i}{\partial t}=\frac{\partial f_i}{\partial T}\frac{\partial T}{\partial t}=\frac{1}{2}\zeta\frac{\partial}{\partial \mathbf{v}}\cdot \left(\mathbf{v}f_i\right),
\eeq
where use has been made of the identity
\beq
\label{3.12.1}
\frac{\partial f_i}{\partial T}=-\frac{1}{2T}\frac{\partial}{\partial \mathbf{v}}\cdot \left(\mathbf{v}f_i\right).
\eeq
Therefore, in dimensionless form, the Enskog equation \eqref{3.1} reads
\beq
\label{3.13}
\frac{1}{2}\zeta_i^* \frac{\partial}{\partial \mathbf{c}}\cdot \left(\mathbf{c}\varphi_i\right)=\sum_{j=1}^s\; \chi_{ij}J_{ij}^*[\mathbf{c}|\varphi_i,\varphi_j],
\eeq
where $\zeta_i^*=\zeta_i/\nu$, $\mathbf{c}=\mathbf{v}/v_\text{th}$, and $J_{ij}^*=(v_\text{th}^d/n_i \nu)J_{ij}^\text{B}$. Here, $\nu(t)=n\overline{\sigma}^{d-1}v_\text{th}(t)$ is an effective collision frequency of the mixture and $\overline{\sigma}=(\sum_i\sigma_i)/s$. The use of $\zeta_i^*$ instead of $\zeta^*=\zeta/\nu$ on the left hand side of Eq.\ \eqref{3.13} is allowed by the equality \eqref{3.11}; this choice is more convenient since the first few velocity moments of Eq.\ \eqref{3.13} are verified without any specification of the distributions $\varphi_i$.

We are in front of a well-possed mathematical problem since we have to solve the set of $s$ Enskog equations \eqref{3.1} for the velocity distribution functions $f_i(v;t)$ of the form \eqref{3.10} and subject to the $s-1$ constraints \eqref{3.11}. These $2s-1$ equations must be solved to determine the $s$ distributions $f_i$ and the $s-1$ temperature ratios $\gamma_i$. As in the case of monocomponent granular gases \cite{NE98}, approximate expressions for the above quantities are obtained by considering the first few terms of the expansion of the distributions $f_i$ in a series of Sonine (or Laguerre) polynomials.

Before obtaining approximate expressions for the temperature ratios, it is important to remark that the failure of energy equipartition in granular fluids has been confirmed in computer simulation works \cite{MG02,BT02,DHGD02,PMP02,CH02,BT02b,KT03,WJM03,BRM05,AL05,SUKSS06,VLSG17,LVGS19,BSG20} and even observed in real experiments of agitated mixtures \cite{WP02,FM02}. All the studies conclude that the departure from energy equipartition depends on the mechanical differences between the particles of the mixture as well as the coefficients of restitution.

\subsection{Approximate solution}

As usual, we expand the distributions $\varphi_i(c)$ in a complete set of orthogonal polynomials with a Gaussian measure. In practice, generalized Laguerre or Sonine polynomials $S_p^{(i)}(c^2)$ are employed. The coefficients $a_p^{(i)}$ of the above expansions are the moments of the distributions $\varphi_i(c)$. These coefficients are obtained by multiplying both sides of the Enskog equation \eqref{3.1} by the polynomials $S_p^{(i)}(c^2)$ and integrating over velocity. It gives an infinite hierarchy for the coefficients $a_p^{(i)}$, which can be approximately solved by retaining only the first few terms of the Sonine polynomial expansion.

The leading Sonine approximation to the distribution $\varphi_i$ is \cite{G19}
\beq
\label{3.14}
\varphi_i(c)=\pi^{-d/2}\theta_i^{d/2} e^{-\theta_i c^2}\Big\{1+\frac{a_2^{(i)}}{2}\Big[\theta_i^2 c^4-(d+2)\theta_i c^2+\frac{d(d+2)}{4}\Big]\Big\},
\eeq
where
\beq
\label{3.15}
\theta_i=\frac{m_i T}{\overline{m} T_i^{(0)}}.
\eeq
Note that the parameters of the Gaussian prefactor in \eqref{3.14} are chosen such that $\varphi_i$ is normalized to 1 and its second moment ($\frac{d}{2}\theta_i^{-1}$) is consistent with the exact moment \eqref{2.8}. An advantage of this choice is that the leading Sonine polynomial is of degree 4. The coefficients $a_2^{(i)}$ measure the departure of $\varphi_i$ from its Maxwellian form $\varphi_{i,\text{M}}=\pi^{-d/2}\theta_i^{d/2} e^{-\theta_i c^2}$. They are defined as
\beq
\label{3.16}
a_2^{(i)}=\frac{4\theta_i^2}{d(d+2)}\int d\mathbf{c}\; c^4 \varphi_i(c)-1.
\eeq
When $\varphi_i=\varphi_{i,\text{M}}$, Eq.\ \eqref{3.16} yields $a_2^{(i)}=0$ as expected. The evaluation of the second Sonine coefficients $a_2^{(i)}$ by considering the contribution to $\varphi_i$ coming from the third Sonine coefficients $a_3^{(i)}$ has been carried out for monocomponent granular gases \cite{BP06a,BP06b}. The results show that the influence of the coefficient $a_3^{(i)}$ on $a_2^{(i)}$ is practically indistinguishable if the (common) coefficient of normal restitution $\al \gtrsim 0.5$. Here, for the sake of simplicity, we will neglect the coefficients $a_3^{(i)}$.

The use of the leading Sonine approximation \eqref{3.14} to $\varphi_i$ permits to estimate the partial cooling rates through their definition \eqref{3.7}. This involves the evaluation of some intricate collision integrals where nonlinear terms in $a_2^{(i)}$ are usually neglected. Such approximation is based on the fact that the coefficients $a_2^{(i)}$ are expected to be very small. In this case, $\zeta_i^*$ can be written as
\beq
\label{3.17}
\zeta_i^*=\zeta_i^{(0)}+\sum_{j=1}^s\; \zeta_{ij}^{(1)}a_2^{(j)},
\eeq
where
\beq
\label{3.18}
\zeta_i^{(0)}=\frac{4\pi^{(d-1)/2}}{d\Gamma\left(\frac{d}{2}\right)}\sum_{j=1}^s\; x_j \chi_{ij}\left(\frac{\sigma_{ij}}{\overline{\sigma}}\right)^{d-1}\mu_{ji}(1+\al_{ij})\left(\frac{\theta_i+\theta_{j}}{\theta_i\theta_j}\right)^{1/2}
\Big[1-\frac{1}{2}\mu_{ji}(1+\al_{ij})\frac{\theta_i+\theta_{j}}{\theta_j}\Big],
\eeq
and the expressions of the quantities $\zeta_{ij}^{(1)}$ are very large to be displayed here. They can be found for instance in Refs.\ \cite{KG14} and \cite{GGG21a}. The temperature ratios $\gamma_i$ can be already determined in the Maxwellian approximation (i.e., when $a_2^{(i)}=0$) by using Eq.\ \eqref{3.18} in the equality of the cooling rates \eqref{3.11}. As will see later, the Maxwellian approximation to $\gamma_i$ leads to a quite accurate predictions.

On the other hand, beyond the Maxwellian approximation, it still remains to estimate the second Sonine coefficients (or kurtosis) $a_2^{(i)}$. To obtain them, we multiply both sides of Eq.\ \eqref{3.13} by $c^4$ and integrate over $\mathbf{c}$. The result is
\beq
\label{3.19}
-\frac{d(d+2)}{2\theta_i^2}\zeta_i^*\left(1+a_2^{(i)}\right)=\sum_{j=1}^s\; \chi_{ij} \int d\mathbf{c}\; c^4\; J_{ij}^*[\varphi_i,\varphi_j]\equiv \Lambda_i.
\eeq
Equation \eqref{3.19} is still exact. However, as in the case of the evaluation of $\zeta_i^*$, the computation of the collision integrals defining $\Lambda_i$ requires the use of the leading Sonine approximation \eqref{3.14} to achieve explicit results. Neglecting nonlinear terms in $a_2^{(i)}$, $\Lambda_i$ can be written as
\beq
\label{3.20}
\Lambda_i=\Lambda_i^{(0)}+\sum_{j=1}^s\; \Lambda_{ij}^{(1)}a_2^{(j)}.
\eeq
The forms of $\Lambda_i^{(0)}$ and $\Lambda_{ij}^{(1)}$ can be found in Refs.\ \cite{KG14} and \cite{GGG21a}. When Eqs.\ \eqref{3.17} and \eqref{3.20} are substituted into Eq.\ \eqref{3.19} and only linear terms in $a_2^{(i)}$ are retained, one gets a system of linear algebraic equations for the coefficients $a_2^{(i)}$:
\beq
\label{3.21}
-\frac{d(d+2)}{2\theta_i^2}\zeta_i^{(0)}-\Lambda_i^{(0)}=\Big[\Lambda_{ii}^{(1)}+\frac{d(d+2)}{2\theta_i^2}
\left(\zeta_i^{(0)}+\zeta_{ii}^{(1)}\right)\Big]a_2^{(i)}+\sum_{j\neq i} \Big[\Lambda_{ij}^{(1)}+\frac{d(d+2)}{2\theta_i^2}\zeta_{ij}^{(1)}\Big]a_2^{(j)}.
\eeq
On the other hand, as noted in several papers on monocomponent granular gases \cite{MS00,CDPT03,SM09}, there is some ambiguity in considering the identity \eqref{3.19} to first order in the coefficients $a_2^{(i)}$. Thus, for instance, if one rewrites Eq.\ \eqref{3.19} as
\beq
\label{3.22}
-\frac{d(d+2)}{2\theta_i^2}\zeta_i^*=\frac{\Lambda_i}{1+a_2^{(i)}},
\eeq
and expands the right hand side as
\beq
\label{3.23}
\frac{\Lambda_i}{1+a_2^{(i)}}\simeq \Lambda_i(1-a_2^{(i)})\simeq \Lambda_i^{(0)}-\Lambda_i^{(0)} a_2^{(i)}+\Lambda_{ii}^{(1)} a_2^{(i)}+\sum_{j\neq i}\Lambda_{ij}^{(1)} a_2^{(j)},
\eeq
one gets the following system of linear algebraic equations:
\beq
\label{3.24}
-\frac{d(d+2)}{2\theta_i^2}\zeta_i^{(0)}-\Lambda_i^{(0)}=\Big[\Lambda_{ii}^{(1)}+\frac{d(d+2)}{2\theta_i^2}\zeta_{ii}^{(1)}-\Lambda_i^{(0)}
\Big]a_2^{(i)}+\sum_{j\neq i} \Big[\Lambda_{ij}^{(1)}+\frac{d(d+2)}{2\theta_i^2}\zeta_{ij}^{(1)}\Big]a_2^{(j)}.
\eeq
The solutions to the set of Eqs.\ \eqref{3.21} and \eqref{3.24} give the second Sonine coefficients $a_2^{(i)}$ as functions of the temperature ratios $\gamma_i$ and the parameters of the mixture (masses and diameters, concentrations, coefficients of restitution, and volume fractions). The accuracy of these solutions will be assessed in Section \ref{sec4} against Monte Carlo simulations in the case of binary mixtures ($s=2$).

When the expressions of the second Sonine coefficients $a_2^{(i)}$ are substituted into the $s-1$ conditions \eqref{3.11} one achieves the $s-1$ temperature ratios $\gamma_i$. The knowledge of $a_2^{(i)}$ and $\gamma_i$ in terms of the parameter space of the system allows us to obtain the scaled distributions $\varphi_i(c)$ in the leading Sonine approximation \eqref{3.14}. This approximate distribution is expected to describe fairly well the behavior of the true distribution in the region of thermal velocities ($v\thicksim v_\text{th}$, say). In the high velocity region (velocities much larger than that of the thermal one), the distributions $\varphi_i$ have an overpopulation [$\varphi_i(c) \thicksim e^{-a c}$] with respect to the Maxwell--Boltzmann tail $e^{-c^2}$ \cite{NE98,EP97,EB02a,EB02b,EB02c,MG02,ETB06}. This exponential decay of the tails of the distribution function has been confirmed by computer simulations \cite{BCR99,HOB00,MG02} and more recently, by means of a microgravity experiment \cite{YSS20}.

To obtain the explicit dependence of the temperature ratios and the Sonine coefficients on the system parameters, the pair correlation functions $\chi_{ij}$ must be given. Although some attempts have been made \cite{L01} for monocomponent granular fluids, we are not aware of any analytical expression of $\chi_{ij}$ for granular mixtures. For this reason, we consider here the approximated expression of $\chi_{ij}$ proposed for molecular mixtures. Thus, in the case of hard spheres ($d=3$), a good approximation for the pair correlation function is \cite{B70,GH72,LL73}
\beq
\label{3.25}
\chi_{ij}=\frac{1}{1-\phi}+\frac{3}{2}\frac{\phi}{(1-\phi)^2}\frac{\sigma_i \sigma_j M_2}{\sigma_{ij}M_3}+\frac{1}{2}\frac{\phi^2}{(1-\phi)^3}
\left(\frac{\sigma_i \sigma_j M_2}{\sigma_{ij}M_3}\right)^2,
\eeq
where $\phi=\sum_i n_i \pi \sigma_i^3/6$ is the solid volume fraction for spheres and $M_\ell=\sum_i x_i \sigma_i^\ell$.


\subsection{Some special limits}

Before illustrating the dependence of the temperature ratios and the second Sonine coefficients on the parameter space for binary ($s=2$) and ternary ($s=3$) mixtures, it is interesting to consider some simple limiting cases. For mechanically equivalent particles ($m_i=m$, $\sigma_i=\sigma$, and $\al_{ij}=\al$), the solution to the conditions \eqref{3.11} yields $\gamma_i=1$ (energy equipartition) while $a_2^{(i)}=a_2$, where
\beq
\label{3.26}
a_2=\frac{16(1-\al)(1-2\al^2)}{9+24d-(41-8d)\al+30(1-\al)\al^2}
\eeq
if we solve Eq.\ \eqref{3.21} or
\beq
\label{3.27}
a_2=\frac{16(1-\al)(1-2\al^2)}{25+24d-(57-8d)\al-2(1-\al)\al^2}
\eeq
if we solve Eq.\ \eqref{3.24}. Equations \eqref{3.26} and \eqref{3.27} agree with the expressions obtained for $a_2$ for monocomponent granular gases \cite{NE98,SM09}, as expected.

Another interesting limit corresponds to the tracer limit, namely, a binary mixture where the concentration of one of the species (for example, species 1) is negligible ($x_1\to 0$). In this limit case, when the collisions between the particles of the excess gas 2 are elastic ($\al_{22}=1$) then the solution to Eqs.\ \eqref{3.21} or \eqref{3.24} lead to $a_2^{(1)}=a_2^{(2)}=0$ and
\beq
\label{3.28}
\frac{T_1^{(0)}}{T_2^{(0)}}=\frac{1+\al_{12}}{2+(1-\al_{12})(m_2/m_1)}.
\eeq
The expression \eqref{3.28} for the temperature ratio agrees with the one derived by Martin and Piasecki \cite{MP99} who found that the Maxwellian distribution with the tracer temperature $T_1^{(0)}$ defined by Eq.\ \eqref{3.28} is an exact solution to the Boltzmann equation in the above conditions ($\al_{22}=1$ and $x_1\to 0$).

We assume now that the tracer particles of the binary mixture are much heavier than particles of the excess gas (Brownian limit, i.e., $m_2/m_1\to 0$). In this limit case, assuming that the temperature ratio $T_1^{(0)}/T_2^{(0)}$ is finite, then the partial cooling rate $\zeta_1$ can be written as
\beq
\label{3.29}
\zeta_1=(1+\al_{12})\left(1-\frac{1+\al_{12}}{2}\frac{T_2^{(0)}}{T_1^{(0)}}\right)
\gamma_e,
\eeq
where
\beq
\label{3.30}
\gamma_e=\frac{4\pi^{(d-1)/2}}{d\Gamma\left(\frac{d}{2}\right)}\chi_{12}n_2
\overline{\sigma}^{d-1}\frac{m_2}{m_1}\sqrt{\frac{2T_2^{(0)}}{m_2}}.
\eeq
The temperature ratio is determined from the solution to the condition $\zeta_2=\zeta_1$. The expresion of $T_1^{(0)}/T_2^{(0)}$ is
\beq
\label{3.31}
\frac{T_1^{(0)}}{T_2^{(0)}}=\frac{1}{2}\left(1+\al_{12}\right)\left(1-\frac{\zeta_2}
{(1+\al_{12})\gamma_e}\right)^{-1},
\eeq
where
\beq
\label{3.32}
\zeta_2=\frac{\sqrt{2}\pi^{(d-1)/2}}{d\Gamma\left(\frac{d}{2}\right)}
(1-\al_{22}^2)
\left(1+\frac{3}{16}a_2\right)n_2\sigma_2^{d-1}\chi_{22}
\sqrt{\frac{2T_2^{(0)}}{m_2}}.
\eeq
Here, depending on the approximation employed, $a_2$ is given by Eqs.\ \eqref{3.26} or \eqref{3.27}. When $\al_{22}=1$, $\zeta_2=0$, and Eq.\ \eqref{3.31} yields
\beq
\label{3.33}
\frac{T_1^{(0)}}{T_2^{(0)}}=\frac{1+\al_{12}}{2}.
\eeq
The expression \eqref{3.33} is consistent with the Brownian limit ($m_2/m_1\to 0$) of Eq.\ \eqref{3.28}. The expressions \eqref{3.29} and \eqref{3.31} agree with the results obtained by Brey \emph{et al.} \cite{BRGD99}. It is important to remark that a ``nonequilibrium'' phase transition \cite{SD01,SD01a} has been found in the Brownian limit which corresponds to a extreme violation of energy equipartition. In other words, there is a region in the parameter space of the system where the temperature ratio goes to infinity and the mean square velocities of the excess gas and the tracer particles remain comparable [$m_1 T_2^{(0)}/(m_2 T_1^{(0)})\equiv \text{finite}$] when the mass ratio $m_1/m_2\to \infty$. Equations \eqref{3.29}--\eqref{3.31} apply of course in the region where $T_1^{(0)}/T_2^{(0)}\equiv \text{finite}$. In this region, the Boltzmann--Lorentz collision operator can be well approximated by the Fokker--Planck operator \cite{BDS99,BP04}.

\section{Comparison between theory and computer simulations}
\label{sec4}

\begin{figure}
\includegraphics[width=0.4\textwidth]{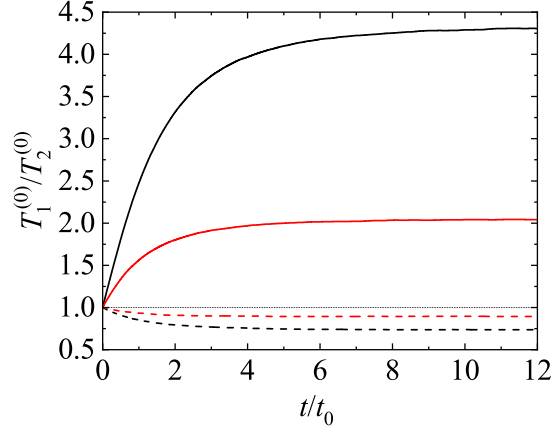}
\caption{Time evolution of the temperature ratio $T_1^{(0)}/T_2^{(0)}$ as obtained from the DSMC method for $d=3$, $\sigma_1/\sigma_2=1$, $x_1=\frac{1}{2}$, $\phi=0$, and two values of the mass ratio: $m_1/m_2=0.5$ (dashed lines) and $m_1/m_2=10$ (solid lines). The black lines refer to $\al_{ij}\equiv \al=0.5$ while the red lines correspond to $\al_{ij}\equiv\al=0.8$. Time is measured in units of $t_0=\ell_{11}/v_{\text{th},1}(0)$.
\label{fig1}}
\end{figure}
\begin{figure}
\includegraphics[width=0.4\textwidth]{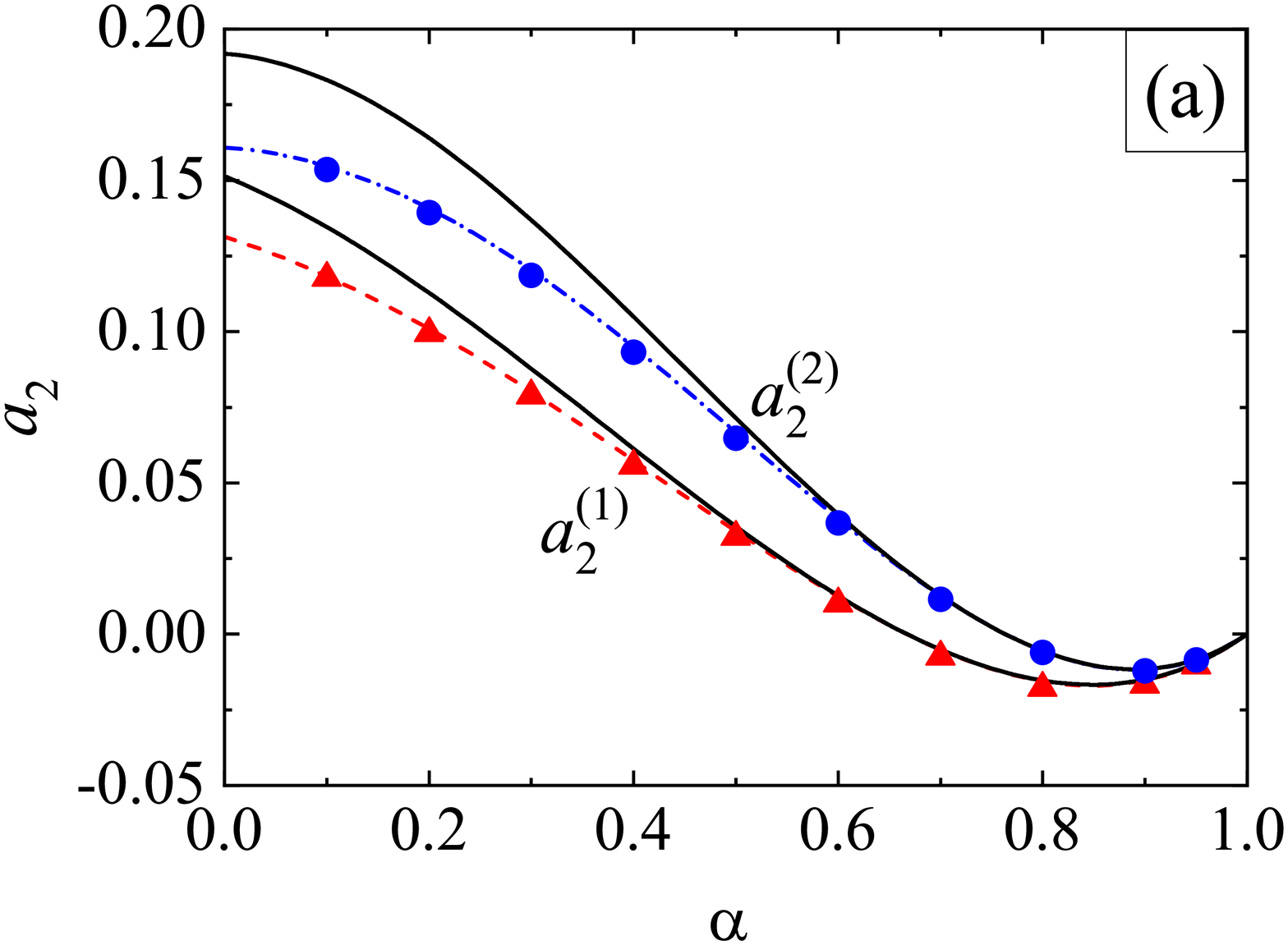}
\includegraphics[width=0.4\textwidth]{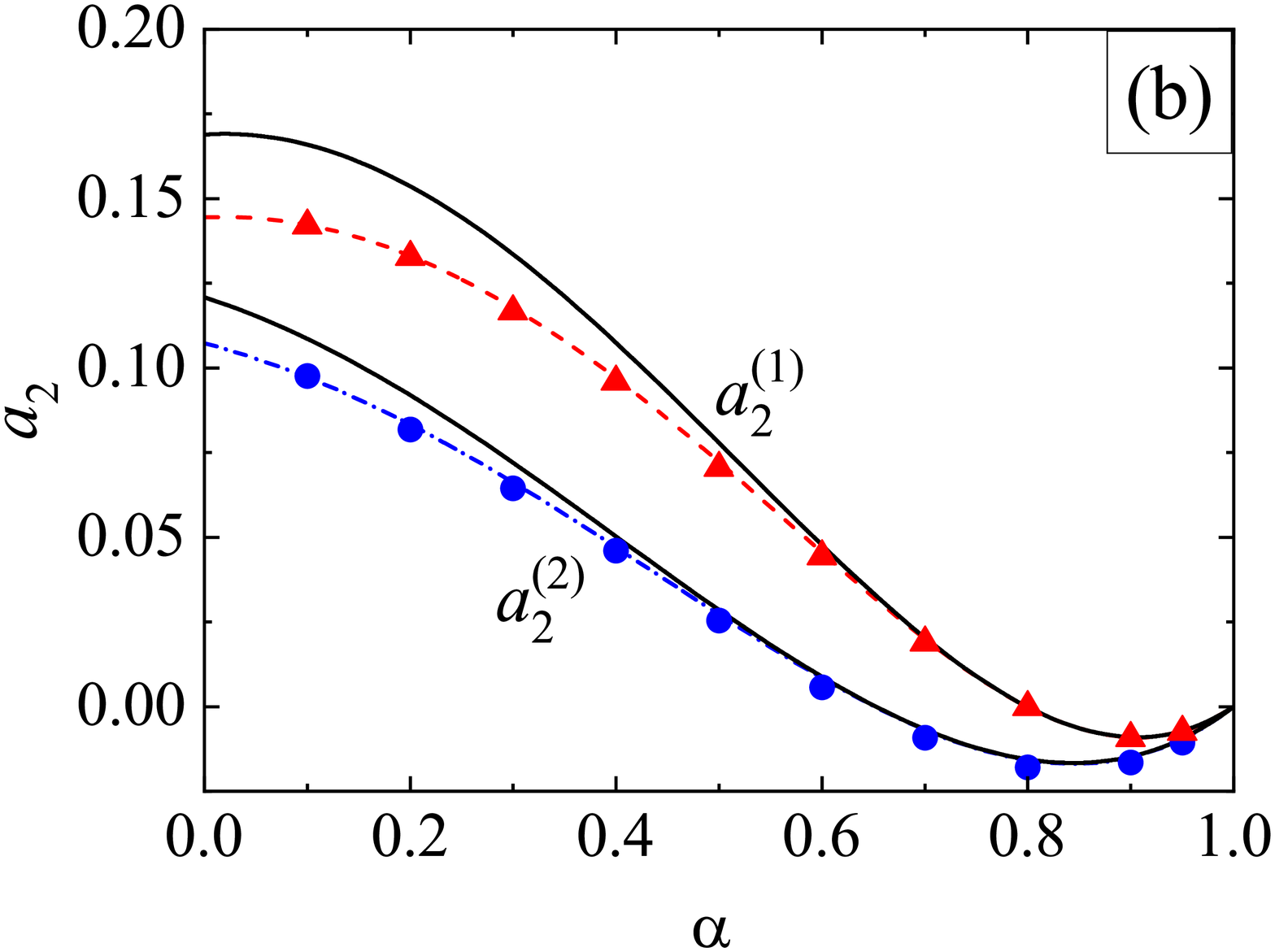}\\ \centering
\includegraphics[width=0.4\textwidth]{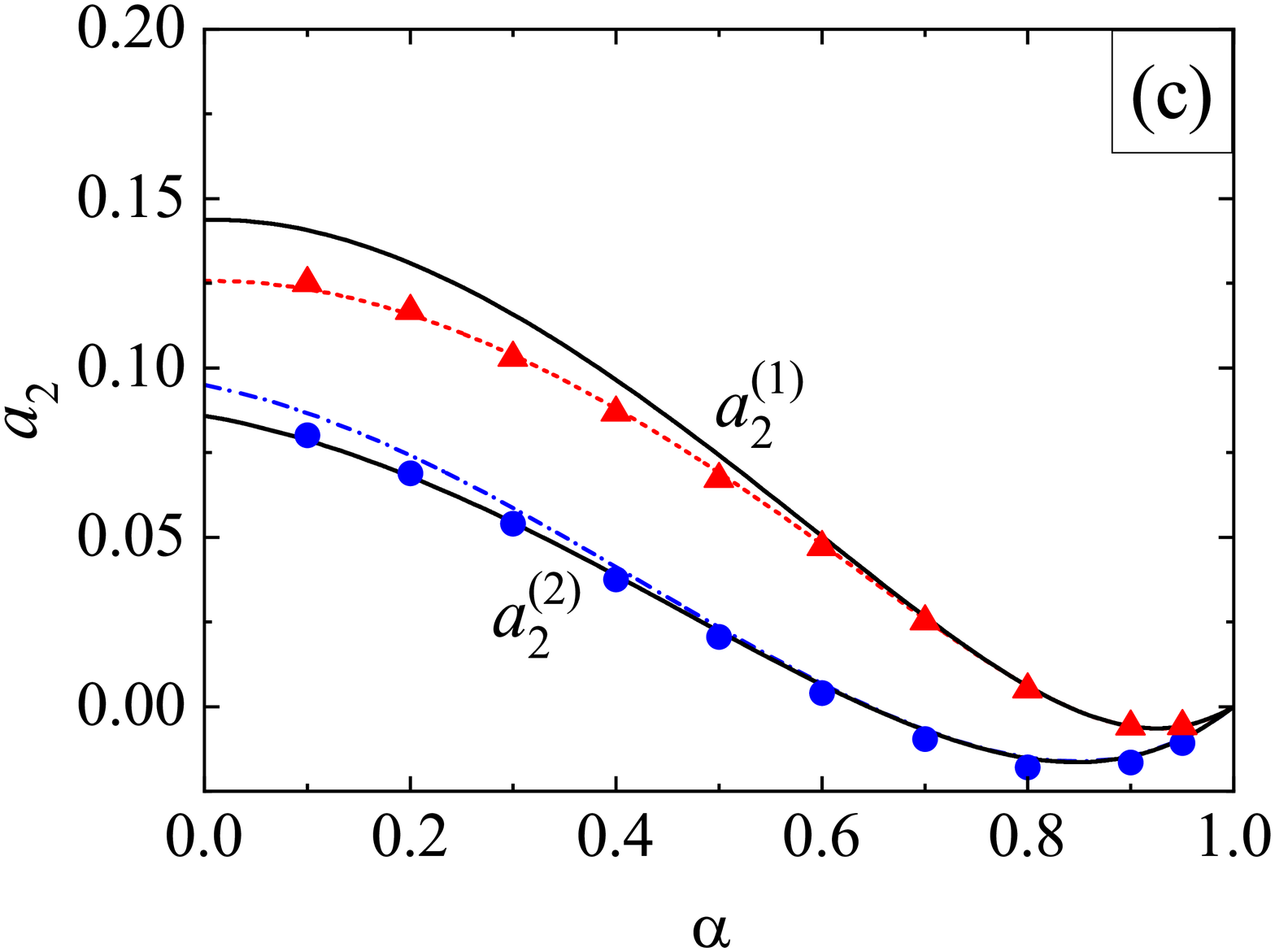}
\caption{Plot of the second Sonine coefficients $a_2^{(1)}$ and $a_2^{(2)}$ versus the (common) coefficient of restitution $\al$ for $\sigma_1/\sigma_2=1$, $x_1=\frac{1}{2}$, $\phi=0$, and three values of the mass ratio: $m_1/m_2=0.5$ (panel (a)), $m_1/m_2=3$ (panel (b)) and $m_1/m_2=5$ (panel (c)). The solid and dashed lines are the theoretical results obtained by solving Eqs.\ \eqref{3.21} and \eqref{3.24}, respectively. Symbols refer to DSMC results (triangles for $a_2^{(1)}$ and circles for $a_2^{(2)})$.
\label{fig2}}
\end{figure}

In the previous section, we have derived expressions for the temperature ratios $\gamma_i$ and the second Sonine coefficients $a_2^{(i)}$ of an $s$-component granular mixture. These (approximate) expressions have been obtained (i) by considering the leading Sonine approximation \eqref{3.14} to the distribution functions and (ii) by retaining only linear terms in $a_2^{(i)}$ in the algebraic equations defining the above coefficients. To asses the degree of accuracy of these theoretical results, in this section we will compare these predictions with those obtained by numerically solving the Enskog equation by means of the well-known DSMC method \cite{B94}. Although this computational method was originally devised for molecular (elastic) fluids, its extension to granular (inelastic) fluids is relatively simple. The simulations allow us to compute the velocity distribution functions over a quite wide range of velocities and obtain precise values of the temperature ratios and the fourth-degree velocity moments in the HCS.
\subsection{DSMC}

In this subsection we provide some details on the application of the DSMC method to a mixture of inelastic hard spheres. More specific details can be found for instance in Ref.\ \cite{MG02}. The DSMC algorithm is composed in its basic form of a collision step that handles all particles collisions and a free drift step between particles collisions. As we are interested in solving the set of \emph{homogeneous} Enskog equations, we take only care of the collisional stage. Thereby, we can consider a single cell wherein the positions of the particles need to be neither computed nor stored.

The velocity distribution function of each species $i$ is represented by the velocities $\{ \mathbf{v_k} \} $ of $N_i$ \emph{simulated} particles:
\begin{equation}
	f_i(\mathbf{v},t)\rightarrow n_i\frac{1}{N_i}\sum_{k=1}^{N_i}\delta(\mathbf{v}-\mathbf{v}_k(t)),
\end{equation}
where $\delta(x)$ is the Dirac delta distribution. The system is initialized by drawing the velocities of the particles from Maxwellian velocity distribution functions with temperatures $T_{i,0}$. Since the system is dilute enough, only binary collisions are considered. Collisions between particles of species $i$ and $j$ are simulated by choosing a sample of $\frac{1}{2}N_i\omega_{\text{max}}^{(ij)}\Delta t$ pairs at random with equiprobability. Here, $\Delta t$ is a time step, which is much smaller than mean free time, and $\omega_{\text{max}}^{(ij)}$ is an upper bound estimate of the probability that a particle of species $i$ collides with a particle of species $j$ per unit of time (typically $\omega_{\text{max}}^{(ij)}=4.0\times v_{th,0}$, where $v_{th,0}=\sqrt{2T(0)/\overline{m}}$ and $T(0)$ is the initial granular temperature). For each pair of particles with velocities $\{ \mathbf{v}_k,\mathbf{v}_\ell \}$ (being $\mathbf{v}_k$ the velocity of a particle of the species $i$ and $\mathbf{v}_\ell$ of the species $j$) a given direction $\bm{\widehat{\sigma}}_{k\ell}$ is chosen at
random with equiprobability.
Then, the collision between particles $i$ and $j$ is accepted with a probability equal to $\bm{\Theta}(\bm{g}_{k\ell} \cdot \bm{\widehat{\sigma}}_{k\ell})\omega_{k\ell}^{(ij)}/\omega_{\text{max}}^{(ij)}$, where $\omega_{k\ell}^{(ij)} = 4\pi\sigma_{ij}^2n_j\chi_{ij}|\bm{g}_{k\ell} \cdot \bm{\widehat{\sigma}}_{k\ell}|$ and $\bm{g}_{k\ell} = \mathbf{v}_k -\mathbf{v}_\ell$. If the collision is accepted, postcollisional velocities of each particle are assigned following the scattering rules \eqref{2.4}. For the cases in which $\omega_{k\ell}^{(ij)}>\omega_{\text{max}}^{(ij)}$, the estimate of $\omega_{\text{max}}^{(ij)}$ is updated as $\omega_{\text{max}}^{(ij)}=\omega_{k\ell}^{(ij)}$. The former procedure is performed for $i=1,2$ and $j=1,2$ in binary and $i=1,2,3$ and $j=1,2,3$ in ternary mixtures.

In the simulations carried out in this work we have typically taken a total number of particles $N=N_1+N_2=10^6$ and five replicas. Since the thermal velocity decreases monotonically with time, we have used a time-dependent time step $\Delta t=3\times10^-3\ell_{11}/v_{th,1}(t)$. Here, $\ell_{11}=(\sqrt{2}\pi n_1 \chi_{11}\sigma_{1}^2)^{-1}$  and $v_{th,1}(t)=\sqrt{2T_1^{(0)}(t)/m_1}$ are the mean free path and the thermal velocity of particles of species 1, respectively.

\subsection{Binary mixtures}

For illustrative purposes, we consider first a binary mixture ($s=2$). The parameter space of this system is constituted by the coefficients of restitution ($\al_{11}$, $\al_{22}$, and $\al_{12}$), the mass ($m_1/m_2$) and diameter ($\sigma_1/\sigma_2$) ratios, the concentration [$x_1=n_1/(n_1+n_2)$], and the solid volume fraction ($\phi$). For the sake of simplicity, henceforth we will consider the case of common coefficients of restitution ($\al\equiv \al_{ij}$) and a three-dimensional system ($d=3$). As discussed before, after an initial transient period, one expects that the scaled distribution functions $\varphi_i(\mathbf{c})$ reach stationary values independent of the initial preparation of the mixture. This hydrodynamic regime is identified as the HCS. In this regime, the temperature ratio $T_1^{(0)}(t)/T_2^{(0)}(t)$ reaches a constant value independent of time. To illustrate the approach toward the HCS, Fig.\ \ref{fig1} shows the time evolution of $T_1^{(0)}(t)/T_2^{(0)}(t)$ obtained from Monte Carlo simulations (DSMC method) for $\sigma_1/\sigma_2=1$, $x_1=\frac{1}{2}$, $\phi=0$, and two values of the mass ratio: $m_1/m_2=0.5$ (dashed lines) and $m_1/m_2=10$ (solid lines). Two coefficients of restitution have been considered: $\al=0.8$ (red lines) and $\al=0.5$ (black lines). Time is measured in units of $t_0=\ell_{11}/v_{\text{th},1}(0)$ where  $v_{\text{th},1}(0)=\sqrt{2T_1^{(0)}(0)/m_1}$, $T_1^{(0)}(0)$ being the initial temperature for species 1. In addition, we have assumed Maxwellian distributions with the same temperature [$T_1^{(0)}(0)=T_2^{(0)}(0)$] at $t=0$. Figure~\ref{fig1} highlights that all the curves converge to different steady values after a relatively short transient period. This clearly confirms the validity of the assumption of constant temperature ratio in the HCS. Although not shown in the graph, the theoretical asymptotic steady values agree very well with their corresponding values obtained from computer simulations.

The second Sonine coefficients $a_2^{(1)}$ and $a_2^{(2)}$ measure the deviations of the scaled distributions $\varphi_1$ and $\varphi_2$, respectively, from their corresponding Maxwellian forms. The panels of Fig.~\ref{fig2} show the $\al$-dependence of the above Sonine coefficients  for $\sigma_1/\sigma_2=1$, $x_1=\frac{1}{2}$, $\phi=0$, and three values of the mass ratio. As for monocomponent granular gases \cite{NE98}, we observe that the coefficients $a_2^{(i)}$ exhibit a non-monotonic dependence on the coefficient of restitution since they decrease first as inelasticity increases until reaching a minimum value and then increase with decreasing $\al$. We also find that the magnitude of $a_2^{(i)}$ is in general very small for not quite strong inelasticity (for instance, $\al \gtrsim 0.5$); this supports the assumption of a low-order truncation in the Sonine polynomial expansion of the distributions $\varphi_i$. With respect to the comparison with computer simulations, it is quite apparent that both theoretical estimates for $a_2^{(i)}$ display an excellent agreement with simulations for values of $\al \gtrsim 0.6$. However, for large inelasticity ($\al\lesssim 0.6$), the best global agreement with simulations is provided by the approach \eqref{3.24}, as Fig.~\ref{fig2} clearly shows for values of $\al \thickapprox 0.1$ (extreme dissipation). Regarding the dependence on the mass ratio, we find that the second Sonine coefficient of the heavier species is larger than that of the lighter species; this means that the departure of $f_1$ from its Maxwellian form accentuates when increasing the mass ratio $m_1/m_2$.

\begin{figure}
\includegraphics[width=0.4\textwidth]{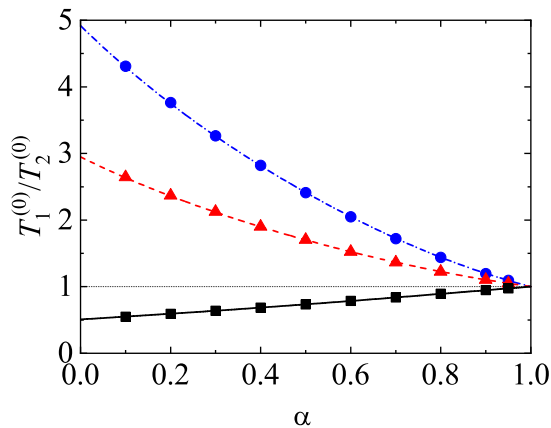}
\caption{Dependence of the temperature ratio $T_1^{(0)}/T_2^{(0)}$ on the coefficient of restitution $\al$  for $\sigma_1/\sigma_2=1$, $x_1=\frac{1}{2}$, $\phi=0$, and three values of the mass ratio: $m_1/m_2=5$ (solid line and squares), $m_1/m_2=3$ (dashed line and triangles), and $m_1/m_2=0.5$ (dash-dotted line and circles). The lines correspond to the theoretical results while symbols refer to the results obtained from the DSMC method.    \label{fig3}}
\end{figure}
\begin{figure}
\includegraphics[width=0.4\textwidth]{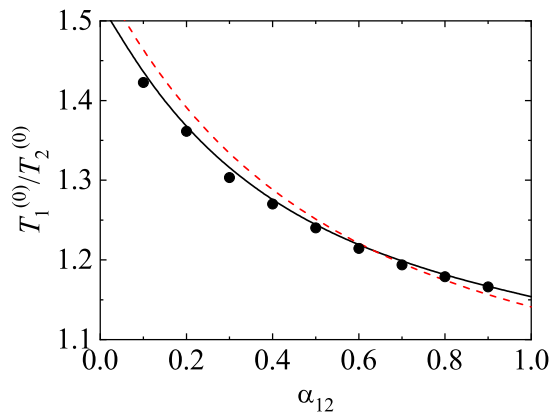}
\caption{Dependence of the temperature ratio $T_1^{(0)}/T_2^{(0)}$ on the coefficient of restitution $\al_{12}$ for $\sigma_1/\sigma_2=1$, $m_1/m_2=1$, $x_1=\frac{1}{2}$, $\phi=0$, $\al_{11}=0.9$, and $\al_{22}=0.5$. The solid line corresponds to the theoretical results obtained by obtaining the second Sonine coefficients $a_2^{(i)}$ by solving Eq.\ \eqref{3.24} while symbols refer to the results obtained from the DSMC method. The dashed line refers to the theoretical prediction in the Maxwellian approximation ($a_2^{(1)}=a_2^{(2)}=0$).}    \label{fig4}
\end{figure}
\begin{figure}
\includegraphics[width=0.4\textwidth]{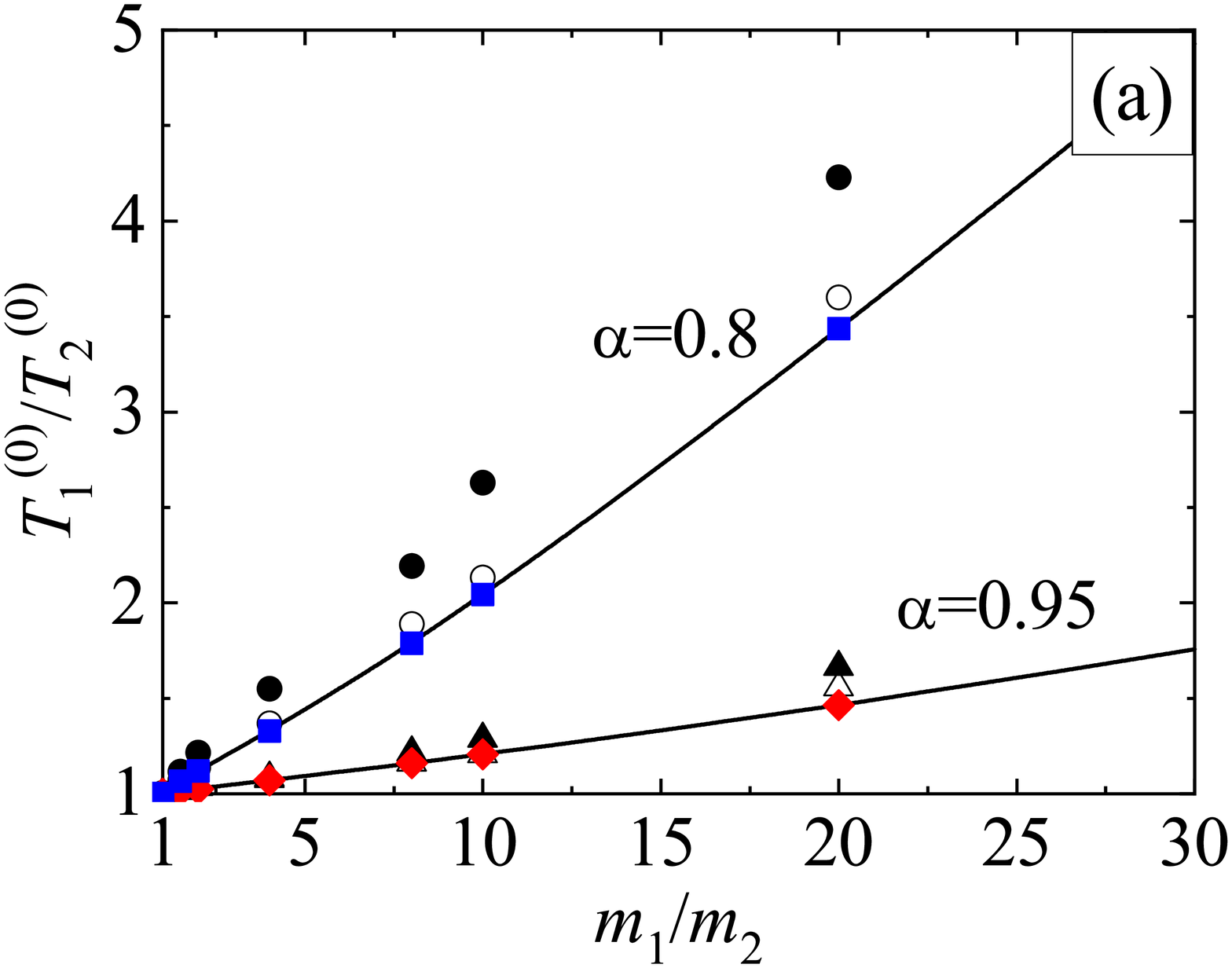}
\includegraphics[width=0.4\textwidth]{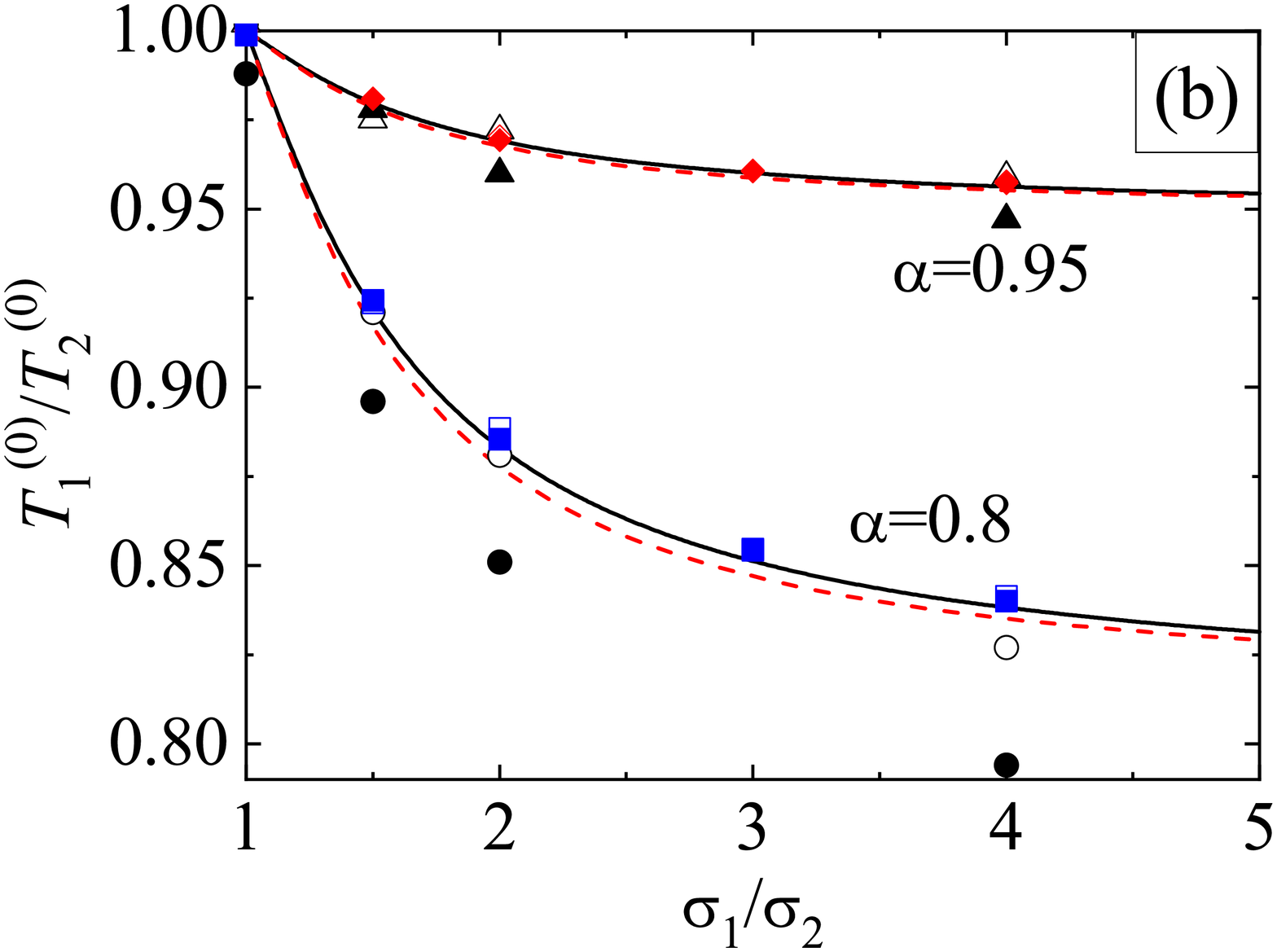}
\caption{Plot of the temperature ratio $T_1^{(0)}/T_2^{(0)}$ as a function of the mass ratio $m_1/m_2$ (panel (a)) and the diameter ratio $\sigma_1/\sigma_2$ (panel (b)) for two different values of the (common) coefficient of restitution $\al$: $\al=0.95$ (lines, triangles and diamonds) and $\al=0.8$ (lines, circles and squares). Triangles and circles refer to MD simulations while diamonds and squares correspond to DSMC results. The open (solid) symbols refer to the volume fraction $\phi=0.1$ ($\phi=0.2$). In the panel (a), $\sigma_1/\sigma_2=1$ and $x_1=\frac{1}{2}$ while in the panel (b), $m_1/m_2=1$ and $x_1=\sigma_2^3/(\sigma_1^3+\sigma_2^3)$. The solid (dashed) lines in the panel (b) correspond to $\phi=0.1$ ($\phi=0.2$).
\label{fig6}}
\end{figure}

One of the most characteristic features of granular mixtures, as compared with molecular mixtures, is that the partial temperatures are different in homogenous states. The breakdown of energy equipartition is clearly illustrated in Fig.~\ref{fig3} where $T_1^{(0)}/T_2^{(0)}$ is plotted versus $\al$ for different mixtures. Here, the solution to Eq.\ \eqref{3.24} is employed to estimate the second Sonine coefficients in the evaluation of the partial cooling rates $\zeta_i$. In any case, the results are practically the same if the solution to Eq.\ \eqref{3.21} for $a_2^{(i)}$ is used. Figure~\ref{fig3} highlights that, at a given value of $\al$, the departure of the temperature ratio from unity increases with increasing the differences in the mass ratio. In general, the temperature of the heavier species is larger than that of the lighter species. Comparison with Monte Carlo simulations shows an excellent agreement in the complete range of values of $\al$. In addition, although not shown here, the theoretical results obtained in the so-called Maxwellian approximation to $\varphi_i$ (i.e, when one takes $a_2^{(1)}=a_2^{(2)}=0$) for $T_1^{(0)}/T_2^{(0)}$ are practically indistinguishable from those derived by considering the second Sonine coefficients. This means that the impact of these coefficients on the partial cooling rates is negligible and so, the Maxwellian approximation to $\varphi_i$ is sufficiently accurate to estimate the temperature ratio.

After having studied the effect of the mass ratio on the temperature ratio, we now turn to further assessing the impact of inelastic collisions on this quantity. To analyze this influence, we consider the case $m_1=m_2$, $\sigma_1=\sigma_2$, but $\al_{11}\neq \al_{22}\neq \al_{12}$. The fact that the coefficients of restitution are different entails that there is breakdown of energy equipartition here either. In other words, the partial temperatures of both species are different when they differ only in their coefficients of restitution. This situation has been widely considered for analyzing segregation driven only by inelasticity \cite{SGNT06,SNTG09,BEGS08,BS09}. The temperature ratio $T_1^{(0)}/T_2^{(0)}$ is plotted versus $\al_{12}$ in Fig.~\ref{fig4} for the above case when $x_1=\frac{1}{2}$, $\phi=0$, $\al_{11}=0.9$, and $\al_{22}=0.5$. The theoretical results have been obtained by solving Eq.\ \eqref{3.24} for getting the coefficients $a_2^{(i)}$ (solid line) and by taking $a_2^{(1)}=a_2^{(2)}=0$ (dashed line). We observe that here the influence of the Sonine coefficients on $T_1^{(0)}/T_2^{(0)}$ is small but not negligible at all since the agreement with simulations improves when these coefficients are considered in the evaluation of the partial cooling rates. We also find that energy nonequipartition is still significant in this particular situation in spite of the fact that the species have the same mass and diameter.

It is quite apparent that although the application of the DSMC method to \emph{dilute} systems is more efficient than the MD method from a computational point of view, the latter method avoids a crucial assumption of the former method: molecular chaos hypothesis (e.g., it neglects the possible velocity correlations between the particles that are about to collide). The study of the HCS of a granular binary mixture from MD simulations allows to prove the existence of the scaled solution \eqref{3.10} in a broader context than the kinetic theory (which is based on molecular chaos assumption). The HCS solution \eqref{3.10} with different partial temperatures determined by equating the partial cooling rates [Eq.\ \eqref{3.11}] has been clearly confirmed by MD simulations \cite{DHGD02}. The occurrence of this sort of solution appears for a wide range of volume fractions, concentrations, and mass and diameter ratios as well as for weak and strong inelasticity. In addition, the comparison between the results obtained from kinetic theory (approximate theoretical results and DSMC results) and MD simulations for the temperature ratio in several conditions may be considered as an stringent assessment of the reliability of kinetic theory. The panels (a) and (b) of Fig.~\ref{fig6} show $T_1^{(0)}/T_2^{(0)}$ versus $m_1/m_2$ and $\sigma_1/\sigma_2$, respectively, for two different values of the coefficient of restitution $\al$. Two different values of the solid volume fraction are considered: $\phi=0.1$ and 0.2 (moderately dense systems). Lines are the approximate theoretical results, circles and triangles refer to MD simulations obtained in Ref.\ \cite{DHGD02} while diamonds and squares correspond to Monte Carlo simulations performed for the present review. The parameters of the granular binary mixture of the panel (a) are $\sigma_1/\sigma_2=1$ and $x_1=\frac{1}{2}$ while in the panel (b) the parameters are $m_1/m_2=1$ and $x_1=\sigma_2^3/(\sigma_1^3+\sigma_2^3)$ (the species volume fraction of each species is the same, i.e., $x_1\sigma_1^3=x_2\sigma_2^3$). Note that altough the systems considered in Fig.~\ref{fig6} correspond to binary mixtures constituted by particles of the same mass [panel (a)] or the same diameter [panel (b)], the theory for the HCS solution applies a priori to arbitary mass or diameter ratios.

Since $\chi_{11}=\chi_{22}=\chi_{12}$ for the parameters chosen in the panel (a) of Fig.~\ref{fig6} [see Eq.\ \eqref{3.25} for $\chi_{ij}$], then the Enskog theoretical predictions are independent of $\phi$. This is confirmed by the DSMC results although MD simulations show a certain dependence of $T_1^{(0)}/T_2^{(0)}$ on $\phi$, specially for $\al=0.8$ and disparate mass ratios ($m_1/m_2=20$). The panel (a) of Fig~\ref{fig6} shows that the agreement between Enskog theory and MD simulations is very good for $\al=0.95$ over the complete range of mass ratios considered. Agreement is also good at $\al=0.8$ and $\phi=0.1$, although significant discrepancies between the Enskog equation (theory and DSMC results) and MD appear for large mass ratios at $\al=0.8$ and $\phi=0.2$. Regarding the dependence of $T_1^{(0)}/T_2^{(0)}$ on $\sigma_1/\sigma_2$, the panel (b) shows a good agreement for both values of the coefficient of restitution at the smallest solid volume fraction $\phi=0.1$, but important differences are observed for the largest solid volume fraction $\phi=0.2$. Thus, both the Enskog theory and Monte Carlo simulations do not accurately predict the value of the temperature ratio found in MD simulations for relatively high densities and/or strong inelasticity.

As widely discussed in molecular mixtures \cite{FK72,DB77}, the Enskog equation has some limitations for describing systems at high densities. In this range of densities, one has to take into account recollision events (ring collisions) which go beyond the Enskog description. The impact of these multiparticle collisions on dynamic properties seems to be more stronger for inelastic collisions due to the fact that colliding pairs tend to be more focused. Thus, one expects that the range of densities where the Enskog equation for granular systems provides reliable predictions diminishes as inelasticity increases \cite{SM01,SPM01}. This is the trend observed here for the temperature ratio and also in other type of problems \cite{LBD02,KG14}. Apart from this limitation, another approximation employed here is the use of Eq.\ \eqref{3.25} for estimating the pair correlation functions $\chi_{ij}$. In particular, recent MD simulations \cite{BSG20} at high densities have shown that $\chi_{11}\neq \chi_{22}\neq \chi_{12}$ even when $x_1=\frac{1}{2}$ and $\sigma_1=\sigma_2$ [the approximation \eqref{3.25} yields $\chi_{11}=\chi_{22}=\chi_{12}$ for this case]. Thus, MD simulations have shown a value of $\chi_{11}$ about a 20\% larger than that of Eq.\ \eqref{3.25}, while $\chi_{22}$ is, however, about 15\% smaller. These differences quantify the effect of the spatial correlations on the Enskog prediction of the temperature ratio.

In conclusion, the comparison carried out in Fig.~\ref{fig6} gives again support to the use of the Enskog equation for the description
of granular flows across a wide range of densities, length scales, and inelasticity. Despite this success, the observed discrepancies between Enskog equation and MD simulations opens the necessity of developing kinetic theories that go beyond the Enskog theory. In any case, as has been
discussed in several previous works \cite{G19}, no other theory with such generality exists yet.

\subsection{Ternary mixtures}

\begin{figure}
\includegraphics[width=0.4\textwidth]{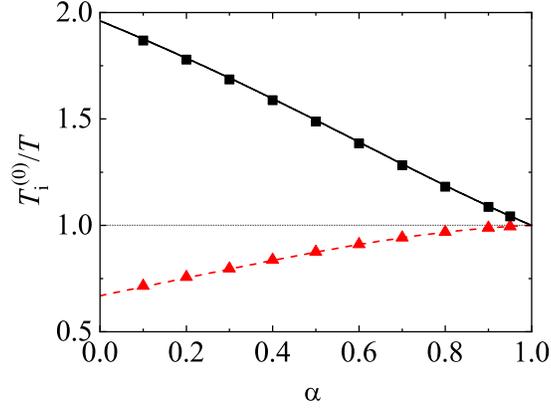}
\caption{Plot of the temperature ratios $T_1^{(0)}/T$ (solid line) and $T_2^{(0)}/T$ (dashed line) versus the (common) coefficient of restitution $\al$ for a dilute ($\phi=0$)} ternary mixture ($s=3$) with  $\sigma_1=\sigma_2=\sigma_3$, $x_1=x_2=\frac{1}{3}$, $m_1/m_3=5$, and $m_2/m_3=2$. Symbols refer to the results obtained from the DSMC method (squares for the case $m_1/m_3=5$ and triangles for the case $m_2/m_3=2$). \label{ternary_dilute}
\end{figure}
\begin{figure}
\includegraphics[width=0.4\textwidth]{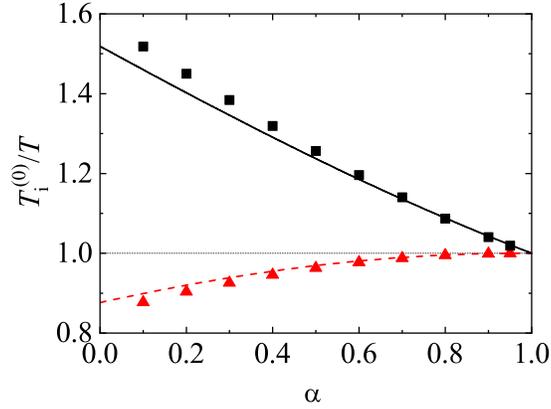}
\caption{Plot of the temperature ratios $T_1^{(0)}/T$ (solid line) and $T_2^{(0)}/T$ (dashed line) versus the (common) coefficient of restitution $\al$ for a ternary mixture ($s=3$) with  $x_1=x_2=\frac{1}{3}$, $\phi=0.1$, $m_1/m_3=5$, $m_2/m_3=2$, $\sigma_1/\sigma_3=(m_1/m_3)^{1/3}$, and $\sigma_2/\sigma_3=(m_2/m_3)^{1/3}$. Symbols refer to the results obtained from the DSMC method (squares for the case $m_1/m_3=5$ and triangles for the case $m_2/m_3=2$).}   \label{ternary_dense}
\end{figure}

Let us consider now a ternary mixture ($s=3$). To the best of our knowledge, the study of this sort of mixtures is scarce in the granular literature \cite{HAM22}. The parameter space in this case is composed by the coefficients of restitution ($\al_{11}$, $\al_{22}$, $\al_{33}$, $\al_{12}$, $\al_{13}$, and $\al_{23}$), the mass ratios ($m_1/m_3$ and $m_2/m_3$), the diameter ratios ($\sigma_1/\sigma_3$ and $\sigma_2/\sigma_3$), the concentrations [$x_1=n_1/(n_1+n_2+n_3)$ and $x_2=n_2/(n_1+n_2+n_3)$], and the solid volume fraction ($\phi$). As in the case of binary mixtures, we assume for simplicity a common coefficient of restitution ($\al_{ij}\equiv \al$) and a three-dimensional system ($d=3$).

Figure \ref{ternary_dilute} shows the $\al$-dependence of the temperature ratios $T_1^{(0)}/T$ and $T_2^{(0)}/T$ for a dilute ($\phi=0$) ternary mixture ($s=3$) with  $\sigma_1=\sigma_2=\sigma_3$, $x_1=x_2=\frac{1}{3}$, $m_1/m_3=5$, and $m_2/m_3=2$. The theoretical results for the temperature ratios have been derived here by neglecting non-Gaussian corrections to the HCS distributions functions ($a_2^{(1)}=a_2^{(2)}=a_2^{(3)}=0$). In spite of this simple approximation, Fig.\ \ref{ternary_dilute} highlights the excellent agreement found between theory and Monte Carlo simulations, even for quite extreme dissipation. As for binary mixtures, the mean kinetic energy of the heavier species is larger than that of the lighter species. Moreover, the departure of the energy equipartition increases with the disparity in the mass ratios.

As a complement of Fig.\ \ref{ternary_dilute}, a moderately dense ternary mixture is considered in Fig.\ \ref{ternary_dense}. Here, $x_1=x_2=\frac{1}{3}$, $\phi=0.1$, $m_1/m_3=5$, $m_2/m_3=2$, $\sigma_1/\sigma_3=(m_1/m_3)^{1/3}$, and $\sigma_2/\sigma_3=(m_2/m_3)^{1/3}$. We observe that the effect of volume fraction on the temperature ratios does not change the main trends observed for dilute ternary mixtures. However, given that the diameter ratios are disparate in this case, more discrepancies between theory and DSMC results are found for small values of the coefficient of restitution, specially when $m_1/m_3=5$. The presence of the second Sonine coefficients in the evaluation of $T_i^{(0)}/T$ could mitigate in part these differences.

\section{Navier--Stokes transport coefficients}
\label{sec5}

We assume that we slightly disturb the HCS by small spatial perturbations. These perturbations induce nonzero contributions to the mass, momentum, and heat fluxes. The corresponding constitutive equations for the irreversible fluxes allow us to identify the relevant Navier--Stokes  transport coefficients of the mixture. As for molecular mixtures, a reliable way of determining the transport coefficients is by means of the Chapman--Enskog method \cite{CC70}. This method solves the set of Enskog equations by expanding the distribution function $f_i(\mathbf{r}, \mathbf{v};t)$ of each species around the \emph{local} version of the HCS (namely, the state obtained from the HCS by replacing the density, flow velocity, and temperature by their local values). The HCS state plays the same role for granular mixtures as the local equilibrium distribution for a molecular mixture (elastic collisions).

Therefore, as in the HCS, after a transient period one assumes that the distributions $f_i$ adopt the form of a \emph{normal} solution. In other words, we assume that all space and time dependence of $f_i(\mathbf{r}, \mathbf{v};t)$ only occurs through a functional dependence on the hydrodynamic fields:
\beq
\label{5.1}
f_i(\mathbf{r}, \mathbf{v};t)=f_i\left[\mathbf{v}|n_i(t), \mathbf{U}(t), T(t)\right].
\eeq
Functional dependence here means that to know $f_i$ at the point $\mathbf{r}$, we need to know the values of the fields and \emph{all} their spatial derivatives at $\mathbf{r}$. For small spatial gradients, the functional dependence \eqref{5.1} can be made local in space through an expansion of $f_i$ in powers of the gradients of the hydrodynamic fields $n_i$, $\mathbf{U}$, and $T$:
\beq
\label{5.2}
f_i(\mathbf{r}, \mathbf{v};t)=f_i^{(0)}(\mathbf{r}, \mathbf{v};t)+f_i^{(1)}(\mathbf{r}, \mathbf{v};t)+\cdots,
\eeq
where the distribution $f_i^{(k)}$ is of order $k$ in gradients. As said before, the reference state $f_i^{(0)}(\mathbf{r}, \mathbf{v};t)$ obeys the Enskog equation \eqref{3.13} but for a global non-homogeneous state (local HCS). The distributions $f_i^{(0)}(\mathbf{r}, \mathbf{v};t)$ are chosen in such a way that their first few velocity moments give the exact hydrodynamic fields:
\beq
\label{5.3}
\int d\mathbf {v} f_i(\mathbf{v})=\int d\mathbf {v} f_i^{(0)}(\mathbf{v})=n_i,
\eeq
\beq
\label{5.4}
\sum_{i=1}^s\int d\mathbf{v}\; m_i \mathbf{v} f_i(\mathbf{v})=\sum_{i=1}^s\int d\mathbf{v}\; m_i \mathbf{v} f_i^{(0)}(\mathbf{v})=\rho \mathbf{U},
\eeq
\beq
\label{5.5}
\sum_{i=1}^s\int d\mathbf{v}\; m_i V^2 f_i(\mathbf{v})=\sum_{i=1}^s\int d\mathbf{v}\; m_i V^2 f_i^{(0)}(\mathbf{v})=d n T.
\eeq
Thus, the remaining distributions $f_i^{(k)}$ must obey the constraints:
\beq
\label{5.6}
\int d\mathbf {v} f_i^{(k)}(\mathbf{v})=0, \quad k\geq 1,
\eeq
\beq
\label{5.7}
\sum_{i=1}^s\int d\mathbf{v}\; \left\{m_i \mathbf{v}, m_i V^2\right\} f_i^{(k)}(\mathbf{v})=\left\{\mathbf{0},0\right\}.
\eeq

It is important to note that in the expansion \eqref{5.2} we have assumed that the spatial gradients are decoupled from the coefficients of restitution. As a consequence, the Navier--Stokes hydrodynamic equations hold for small spatial gradients but they are not limited in principle to weak inelasticity. This point is relevant in the case of granular mixtures since there are some situations (e.g., steady states such as the uniform shear flow problem \cite{JR88,MG02a,SGD04,VSG10,VGS11}) where hydrodynamic gradients are coupled to inelasticity and so, the Navier--Stokes approximation is restricted to nearly elastic spheres. Thus, due to the possible lack of scale separation for strong inelasticity, Serero \emph{et al.} \cite{SGNT06,SNTG09} consider two different perturbation parameters in the Chapman--Enskog solution: the hydrodynamic gradients (or equivalently, the Knudsen number $\text{Kn}=\ell/L$, where $\ell$ is the mean free path and $L$ is a characteristic hydrodynamic length) and the degree of dissipation $\epsilon_{ij}=1-\al_{ij}^2$. The results derived from this perturbation scheme \cite{SGNT06,SNTG09} agree with those obtained here in the quasielastic limit ($\epsilon_{ij}\to 0$).

Another important issue in the Chapman--Enskog expansion of granular mixtures is the choice of the hydrodynamic fields. Here, as for molecular mixtures \cite{CC70,LCK83,KLC83,LC84,KCL87}, we use the conserved number densities $n_i$, the flow velocity $\mathbf{U}$ associated with the conserved total momentum, and the granular temperature $T$ associated with the total kinetic energy. On the other hand, due to energy nonequipartition, other authors \cite{HWRLG00,HGM01,RNW03,ChMW17,RSN20,SM21a,SM21b,ZW21} employ the set consisting of the conserved number densities $n_i$, the species flow velocities $\mathbf{U}_i$ associated with the non-conserved species momenta, and the partial (or species) temperatures $T_i$. However, this choice is potentially confusing since, although more detailed, has no predictive value on the relevant hydrodynamic large space and time scales \cite{DB11}. In particular, the two-temperature Chapman--Enskog solution considered in these works \cite{HWRLG00,HGM01,RNW03,ChMW17,RSN20,SM21a,SM21b,ZW21} is phenomenological and assumes local Maxwellian distributions even for non-homogeneous situations. Although this approach yields vanishing Navier--Stokes transport coefficients for low-density mixtures, it can be considered as reliable to estimate the collisional transfer contributions to the irreversible fluxes \cite{SM21b,G21}.

The Chapman--Enskog solution to the (inelastic) Enskog equation \eqref{2.1} was obtained in Refs.\ \cite{GDH07,GHD07} some years ago. In particular, to first order in spatial gradients, the first-order velocity distribution function $f_i^{(1)}(\mathbf{r}, \mathbf{v};t)$ is
\beq
\label{5.8}
f_i^{(1)}=\boldsymbol{\mathcal{A}}_i\cdot\nabla\ln T+\sum_{j=1}^s\boldsymbol{\mathcal{B}}_{ij}\cdot \nabla\ln n_j+ \mathcal{C}_{i,\lambda\beta}\frac{1}{2}\left(\frac{\partial U_\beta}{\partial r_\lambda}+\frac{\partial U_\lambda}{\partial r_\beta}-\frac{2}{d}\delta_{\lambda\beta}\nabla\cdot\mathbf{U}\right)+\mathcal{D}_i\nabla\cdot\mathbf{U},
\eeq
where the unknowns $\boldsymbol{\mathcal{A}}_i(\mathbf{V})$, $\boldsymbol{\mathcal{B}}_{ij}(\mathbf{V})$, $\mathcal{C}_{i,\lambda\beta}(\mathbf{V})$, and $\mathcal{D}_i(\mathbf{V})$ are functions of the peculiar velocity $\mathbf{V}$. These quantities are the solutions of a set of coupled linear integral equations \cite{GDH07}. Approximate solutions to this set of integral equations were obtained \cite{GHD07,MGH12} by considering the leading terms in a Sonine polynomial expansion. This procedure allows us to get the explicit forms of the Navier--Stokes transport coefficients in terms of the mechanical parameters of the mixture (masses and sizes and the coefficients of restitution), the composition, and the solid volume fraction.

The constitutive equations for the mass $\mathbf{j}_i^{(1)}$, momentum $P_{\lambda\beta}^{(1)}$, and heat $\mathbf{q}^{(1)}$ fluxes have the form
\beq
\label{5.9}
\mathbf{j}_i^{(1)}=-\sum_{j=1}^s \frac{m_i m_j n_j}{\rho} \; D_{ij}\;  \nabla \ln n_j-\rho\; D_i^T \; \nabla \ln T,
\eeq
\beq
\label{5.10}
P_{\lambda\beta}^{(1)}=-\eta \left(\frac{\partial U_\beta}{\partial r_\lambda}+\frac{\partial U_\lambda}{\partial r_\beta}-\frac{2}{d}\delta_{\lambda\beta}\nabla\cdot\mathbf{U}\right) -\delta_{\lambda\beta}\eta_\text{b} \nabla \cdot \mathbf{U},
\eeq
\beq
\label{5.11}
\mathbf{q}^{(1)}=-\sum_{i=1}^s\sum_{j=1}^s T^2\; D_{q,ij} \nabla \ln n_j-T \; \kappa \; \nabla \ln T.
\eeq
In Eqs.\ \eqref{5.9}--\eqref{5.11}, $D_{ij}$ are the mutual diffusion coefficients, $D_i^T$ are the thermal diffusion coefficients, $\eta$ is the shear viscosity coefficient, $\eta_\text{b}$ is the bulk viscosity, $\kappa$ is the thermal conductivity coefficient, and $D_{q,ij}$ are the partial contributions to the Dufour coefficients $D_{q,i}=\sum_j D_{q,ji}$.

The Navier--Stokes transport coefficients associated with the mass flux are defined as
\beq
\label{5.12}
D_{ij}=-\frac{\rho}{d m_j n_j}\int d\mathbf{v}\; \mathbf{V}\cdot \boldsymbol{\mathcal{B}}_{ij}(\mathbf{V}),
\eeq
\beq
\label{5.13}
D_{i}^T=-\frac{m_i}{d\rho}\int d\mathbf{v}\; \mathbf{V}\cdot \boldsymbol{\mathcal{A}}_{i}(\mathbf{V}).
\eeq
The Navier--Stokes transport coefficients associated with the pressure tensor and the heat flux have kinetic and collisional contributions. Their kinetic contributions are given by $\eta_\text{b}^k=0$,
\beq
\label{5.14}
\eta_k=\sum_{i=1}^s\; \eta_i^k, \quad \eta_i^k=-\frac{1}{(d+2)(d-1)} \int d\mathbf{v}\; m_i V_\lambda V_\beta \mathcal{C}_{i,\lambda\beta}(\mathbf{V}),
\eeq
\beq
\label{5.15}
D_{q,ij}^k=-\frac{1}{d T^2}\int d\mathbf{v}\; \frac{m_i}{2}V^2\mathbf{V}\cdot \boldsymbol{\mathcal{B}}_{ij}(\mathbf{V}),
\eeq
\beq
\label{5.16}
\kappa_k=\sum_{i=1}^s\; \kappa_i^k, \quad \kappa_i^k=-\frac{1}{d} \int d\mathbf{v}\; \frac{m_i}{2}V^2\mathbf{V}\cdot \boldsymbol{\mathcal{A}}_{i}(\mathbf{V}).
\eeq
The expressions of the collisional contributions to $\eta$, $\eta_\text{b}$, $D_{q,ij}$, and $\kappa$ can be determined from Eqs.\ \eqref{2.19} and \eqref{2.21} by expanding the distribution functions $f_i$ to first order in gradients. Their explicit forms can be found in Ref.\ \cite{G19}. We will go back to this point in section \ref{sec7} when we analyze the impact of different partial temperatures on the bulk viscosity coefficient.

\section{Influence of the temperature ratios $T_i^{(0)}/T$ on the transport coefficients}
\label{sec6}

As mentioned before, the determination of the set of Navier--Stokes transport coefficients requires to know the functions $\boldsymbol{\mathcal{A}}_i(\mathbf{V})$, $\boldsymbol{\mathcal{B}}_{ij}(\mathbf{V})$, $\mathcal{C}_{i,\lambda\beta}(\mathbf{V})$, and $\mathcal{D}_i(\mathbf{V})$. As in the study of the HCS, the usual approach is to expand these unknowns in a series expansion of Sonine polynomials and consider only the leading terms. This procedure involves a quite long and tedious task where several collision integrals must be computed.

As expected, all the transport coefficients depend explicitly on the temperature ratios $\gamma_i=T_i^{(0)}/T$, which are defined in terms of the zeroth-order distributions $f_i^{(0)}$. As discussed in section \ref{sec3}, given that the form of $f_i^{(0)}(\mathbf{V})$ is not exactly known, one considers the leading Sonine approximation \eqref{3.14} (namely, a polynomial in velocity of degree four) to the scaled distribution $\varphi_i$. However, the results obtained in section \ref{sec4} for the HCS have clearly shown that the effect of the second Sonine coefficients $a_2^{(i)}$ on the temperature ratios $\gamma_i$ is very tiny. Thus, for practical purposes, one can replace $f_i^{(0)}(\mathbf{V})$ by the Maxwellian distribution
\beq
\label{6.1}
f_{i,\text{M}}(\mathbf{V})=n_i \left(\frac{m_i}{2\pi T_i^{(0)}}\right)^{d/2} \exp \left(-\frac{m_i V^2}{2T_i^{(0)}}\right).
\eeq
In this approximation, the (reduced) partial cooling rates $\zeta_i^*\to \zeta_i^{(0)}$, where $\zeta_i^{(0)}$ is given by Eq.\ \eqref{3.18}.

The forms of the Navier--Stokes transport coefficients can be found in Ref.\ \cite{G19} when one uses the Maxwellian approximation \eqref{6.1}. Their explicit expressions are very large and so, they are omitted here for the sake of brevity. On the other hand, for the sake of concreteness and to show in a clean way the impact of $\gamma_i$ on transport, we focus on our attention in this section in the coefficients $D_{ij}$ and $D_i^T$ of a binary mixture ($s=2$) in the low-density regime ($\phi=0$). The diffusion coefficients play for instance a relevant role in one of the most important applications in granular mixtures: segregation by thermal diffusion \cite{G11}. Since $\mathbf{j}_1^{(1)}=-\mathbf{j}_2^{(1)}$ for $s=2$, then one has the relations $D_{21}=-(m_2/m_1)D_{11}$, $D_{22}=-(m_2/m_1)D_{12}$, and $D_2^T=-D_1^T$. In dimensionless form, these coefficients can be written as \cite{G19}
\beq
\label{6.2}
D_{ij}=\frac{\rho T}{m_i m_j \nu}D_{ij}^*, \quad D_1^T=\frac{n T}{\rho \nu}D_1^{T*},
\eeq
where
\beq
\label{6.3}
D_1^{T*}=\frac{x_1}{\nu_D^*-\zeta_0^*}\left(\gamma_1-\frac{m_1 n}{\rho}\right),
\eeq
\beq
\label{6.4}
D_{11}^*=\Big(\nu_D^*-\frac{1}{2}\zeta_0^*\Big)^{-1}\Bigg[\Big(\zeta_0^*+x_2\frac{\partial \zeta_0^*}{\partial x_1}\Big)D_1^{T*}-\frac{\rho_1}{\rho}+\gamma_1+x_1 x_2 \frac{\partial \gamma_1}{\partial x_1}\Bigg],
\eeq
\beq
\label{6.5}
D_{12}^*=\Big(\nu_D^*-\frac{1}{2}\zeta_0^*\Big)^{-1}\Bigg[\Big(\zeta_0^*-x_1\frac{\partial \zeta_0^*}{\partial x_1}\Big)D_1^{T*}-\frac{\rho_1}{\rho}-x_1^2 \frac{\partial \gamma_1}{\partial x_1}\Bigg].
\eeq
Here,
\beq
\label{6.6}
\nu_D^*=\frac{2\pi^{(d-1)/2}}{d\Gamma\left(\frac{d}{2}\right)}\left(\frac{\theta_1+\theta_2}
{\theta_1\theta_2}\right)^{1/2}
\left(x_1\mu_{12}+x_2\mu_{21}\right)(1+\al_{12}),
\eeq
$\zeta_0^*=\zeta^{(0)}/\nu$, where $\zeta^{(0)}=\zeta_1^{(0)}=\zeta_2^{(0)}$ is given by Eq.\ \eqref{3.18} with $\chi_{ij}=1$.

It is quite apparent from Eqs.\ \eqref{6.4}--\eqref{6.6} that the coefficients $D_{ij}^*$ and $D_1^{T*}$ depend in a complex way on the temperature ratio $\gamma_1$ [recall that  $\gamma_2=x_2^{-1}(1-x_1 \gamma_1)$ in a binary mixture]. To show more clearly the influence of energy nonequipartition on diffusion coefficients, it is convenient to write the forms of the above dimensionless coefficients by assuming energy equipartition. In this approximation ($\gamma_1=1$), $\theta_1=2\mu_{12}$, $\theta_2=2\mu_{21}$,
\beq
\label{6.7}
\zeta_0^*+x_2\frac{\partial \zeta_0^*}{\partial x_1}\equiv \overline{\zeta}_{11}, \quad \overline{\zeta}_{11}=\frac{\pi^{(d-1)/2}}{d\Gamma\left(\frac{d}{2}\right)}
\left(\frac{\sigma_1}{\sigma_{12}}\right)^{d-1}\left(\frac{m_1+m_2}{m_1}\right)^{1/2}(1-\al_{11}^2),
\eeq
\beq
\label{6.8}
\zeta_0^*-x_1\frac{\partial \zeta_0^*}{\partial x_1}\equiv \overline{\zeta}_{12}, \quad \overline{\zeta}_{12}=\frac{\sqrt{2}\pi^{(d-1)/2}}{d\Gamma\left(\frac{d}{2}\right)}\left(\frac{m_2}{m_1}\right)^{1/2}(1-\al_{12}^2),
\eeq
$\zeta_0^*\equiv \zeta_{0,\text{eq}}^*=x_1\overline{\zeta}_{11}
+x_2\overline{\zeta}_{12}$, and
\beq
\label{6.9}
\nu_D^*\equiv \nu_{D,\text{eq}}^*=\frac{\sqrt{2}\pi^{(d-1)/2}}{d\Gamma\left(\frac{d}{2}\right)}
\frac{\rho}{n\sqrt{m_1 m_2}}(1+\al_{12}).
\eeq
Here, we recall that $\sigma_{12}=(\sigma_1+\sigma_2)/2$ and $\rho=m_1 n_1+m_2 n_2$. Thus, taking into account Eqs.\ \eqref{6.7}--\eqref{6.9}, the forms of $D_{ij}^*$ and $D_1^{T*}$ by assuming energy equipartition read
\beq
\label{6.10}
D_{1,\text{eq}}^{*T}=\frac{x_1}{\nu_{D,\text{eq}}^*-
\zeta_{0,\text{eq}}^*}\frac{n_2(m_2-m_1)}{\rho},
\eeq
\beq
\label{6.11}
D_{11,\text{eq}}^*=\frac{
\overline{\zeta}_{11}D_{1,\text{eq}}^{*T}+\frac{\rho_2}{\rho}}{\nu_{D,\text{eq}}^*-\frac{1}{2}\zeta_{0,\text{eq}}^*},\quad
D_{12,\text{eq}}^*=
\frac{\overline{\zeta}_{12}D_{1,\text{eq}}^{*T}-\frac{\rho_1}{\rho}}
{\nu_{D,\text{eq}}^*-\frac{1}{2}\zeta_{0,\text{eq}}^*}.
\eeq

\begin{figure}
\includegraphics[width=0.4\textwidth]{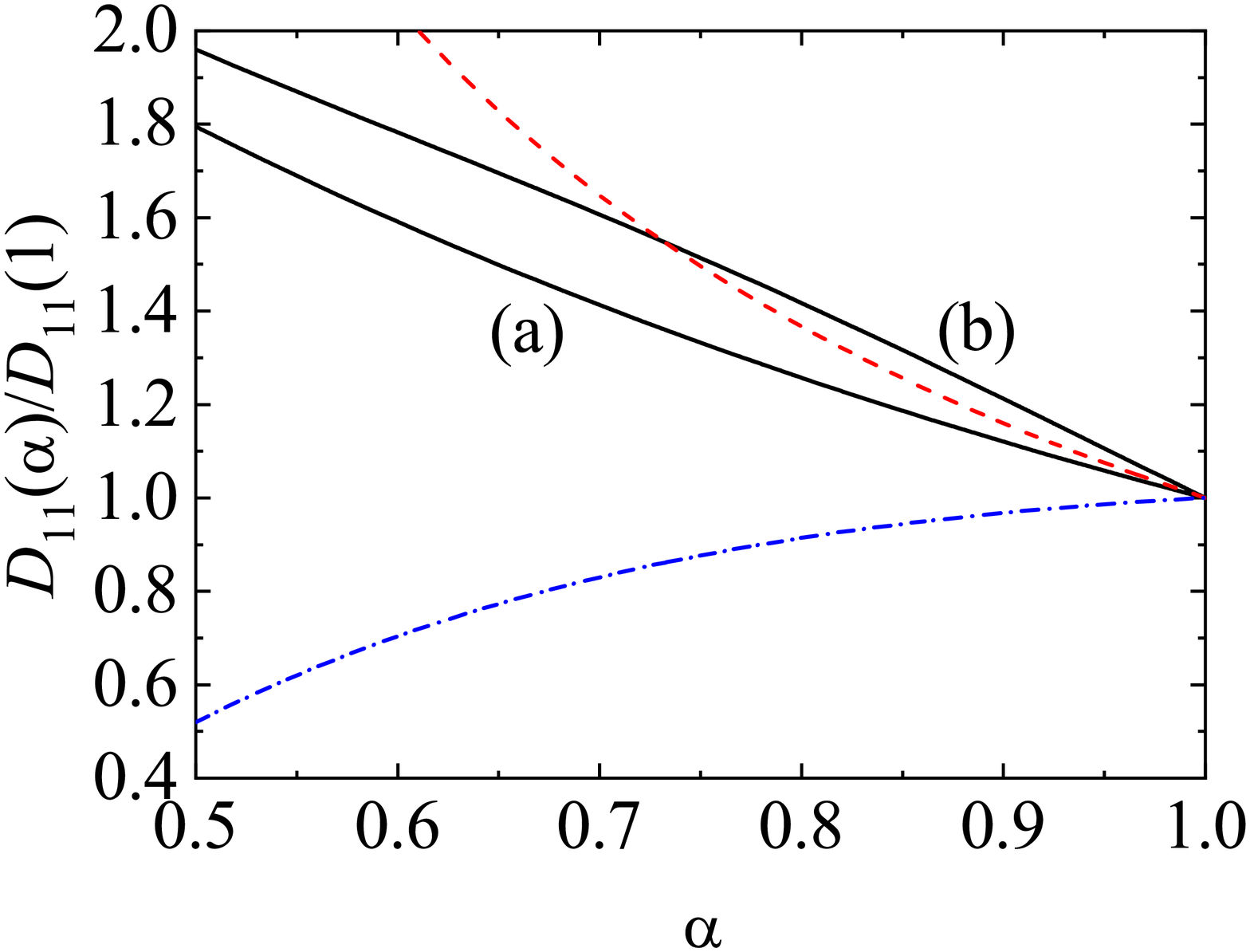}
\includegraphics[width=0.4\textwidth]{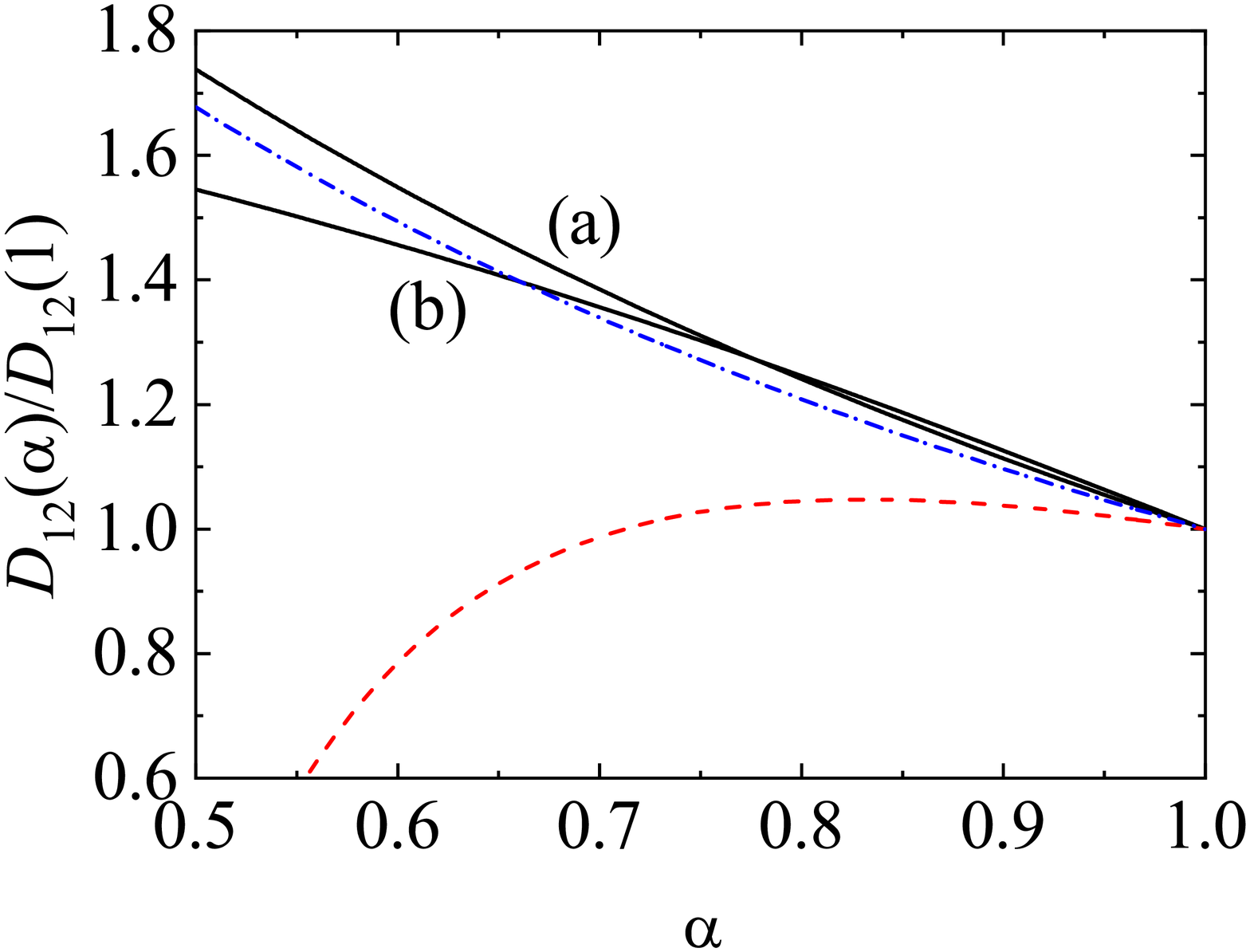}\\ \centering
\includegraphics[width=0.4\textwidth]{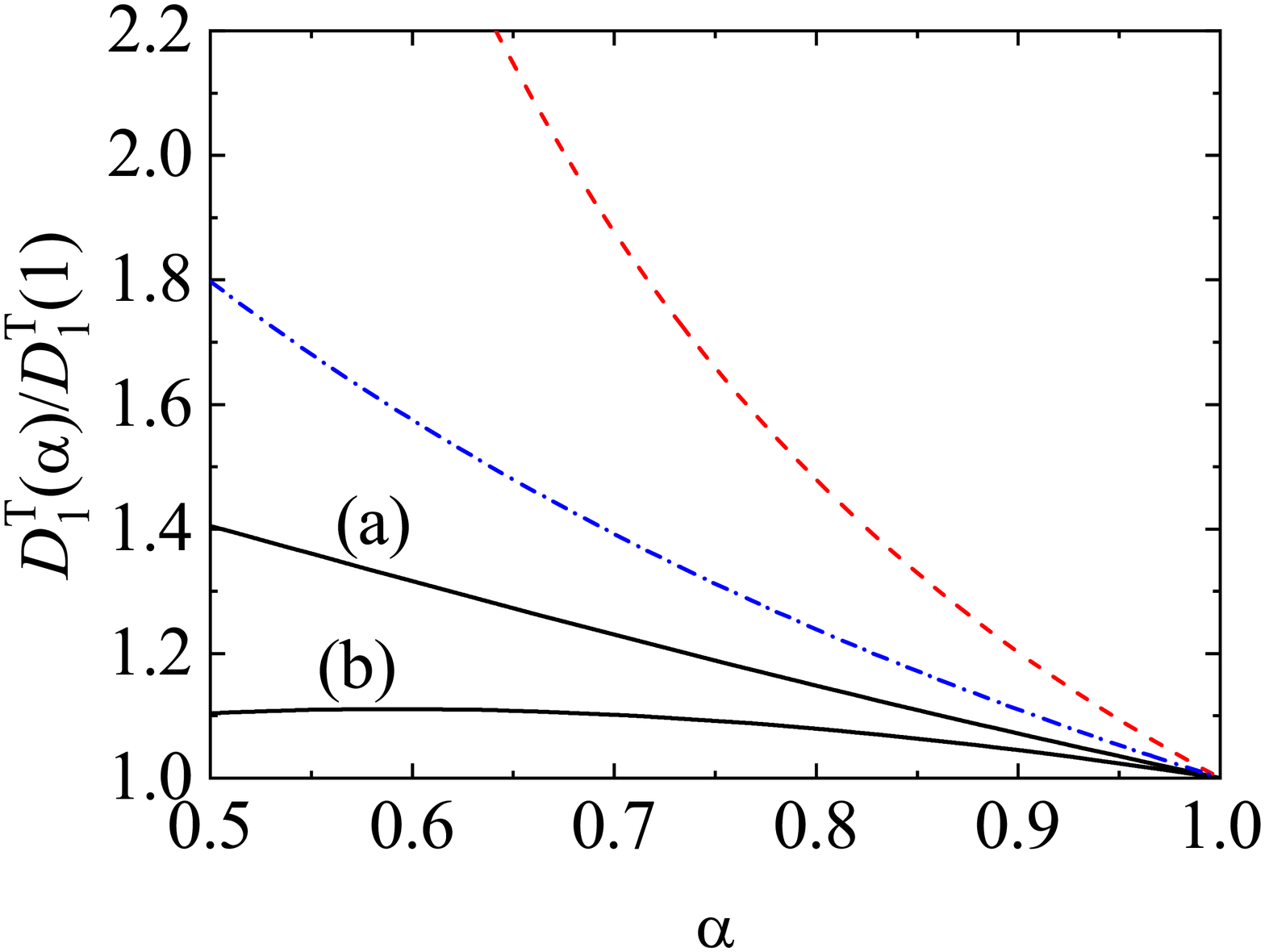}
\caption{Dependence of the (scaled) diffusion transport coefficients $D_{11}(\al)/D_{11}(1)$, $D_{12}(\al)/D_{12}(1)$, and $D_1^T(\al)/D_1^T(1)$ on the coefficient of restitution $\al$ for a three-dimensional ($d=3$) binary mixture ($s=2$) with $\sigma_1=\sigma_2$, $x_1=\frac{1}{2}$, $\phi=0$, and two different values of the mass ratio: $m_1/m_2=0.5$ (a) and $m_1/m_2=4$ (b). The dashed and dash-dotted lines correspond to the cases $m_1/m_2=0.5$ and $m_1/m_2=4$, respectively, by assuming energy equipartition [Eqs.\ \eqref{6.10}--\eqref{6.11}].
\label{diffusion}}
\end{figure}
\begin{figure}
\includegraphics[width=0.4\textwidth]{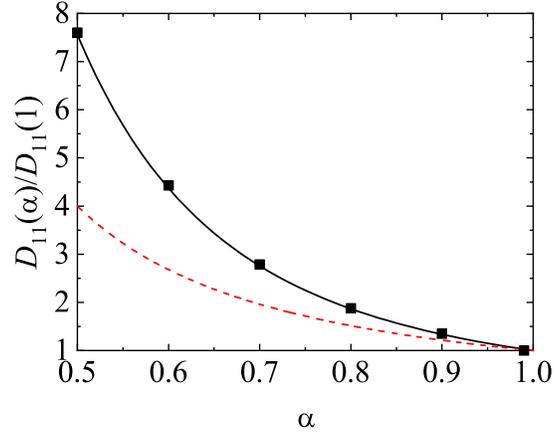}
\caption{Dependence of the (scaled) tracer diffusion coefficient $D_{11}(\al)/D_{11}(1)$ on the (common) coefficient of restitution $\al$ for a dilute ($\phi=0$) three-dimensional ($d=3$) binary mixture ($s=2$) in the tracer limit ($x_1\to 0$) with $\sigma_1/\sigma_2=2$, and $m_1/m_2=8$. The solid line corresponds to the theoretical result obtained from Eq.\ \eqref{6.11.1} and the dashed line refers to the result obtained from Eq.\ \eqref{6.11.1} but assuming energy equipartition. Symbols refer to the DSMC results reported in Ref.\ \cite{GM04}.
\label{tracer}}
\end{figure}

\begin{figure}
\includegraphics[width=0.4\textwidth]{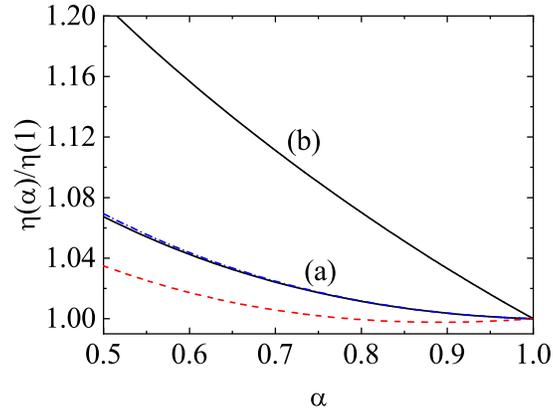}
\caption{Dependence of the (scaled) shear viscosity coefficient $\eta(\al)/\eta(1)$ on the (common) coefficient of restitution $\al$ for a three-dimensional ($d=3$) binary mixture ($s=2$) with $\sigma_1=\sigma_2$, $x_1=0.5$, $\phi=0.1$, and two different values of the mass ratio: $m_1/m_2=0.5$ (a) and $m_1/m_2=4$ (b). The dashed and dash-dotted lines correspond to the cases $m_1/m_2=0.5$ and $m_1/m_2=4$, respectively, by assuming energy equipartition.
\label{viscosity}}
\end{figure}
\begin{figure}
\includegraphics[width=0.4\textwidth]{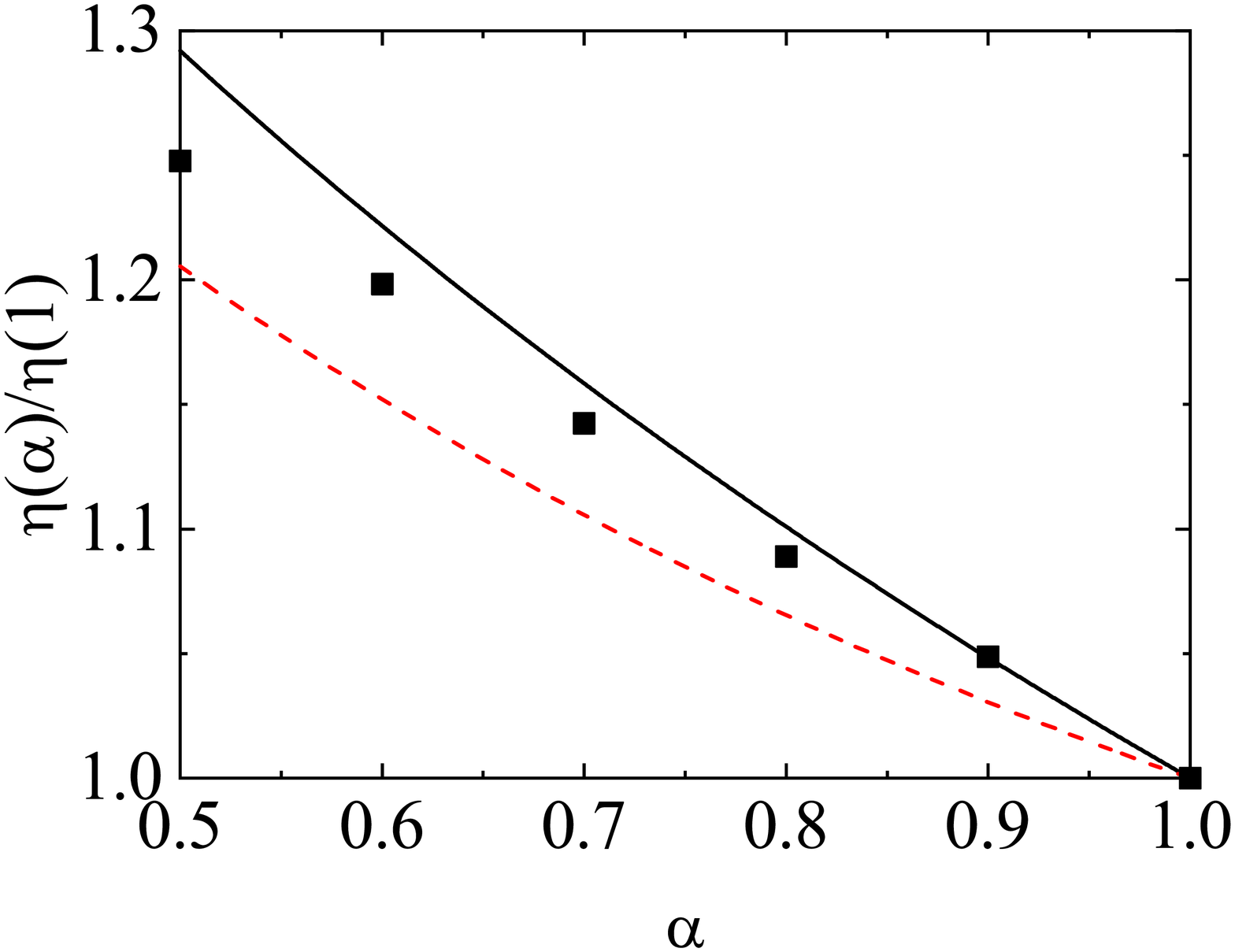}
\caption{Dependence of the (scaled) shear viscosity coefficient $\eta(\al)/\eta(1)$ on the (common) coefficient of restitution $\al$ for a dilute ($\phi=0$) two-dimensional ($d=2$) binary mixture ($s=2$) constituted by particles of the same mass density [i.e., $m_1/m_2=(\sigma_1/\sigma_2)^2$]. The dashed line corresponds to the result obtained by assuming energy equipartition. Symbols refer to the simulation results obtained from the DSMC method in Ref.\ \cite{GM07}.   \label{viscosity2d}}
\end{figure}

Figure~\ref{diffusion} shows the scaled diffusion coefficients $D_{11}(\al)/D_{11}(1)$, $D_{12}(\al)/D_{12}(1)$, and $D_1^T(\al)/D_1^T(1)$ versus the (common) coefficient of restitution $\al_{11}=\al_{22}=\al_{12}\equiv \al$ for a three-dimensional dilute binary mixture with $\sigma_1/\sigma_2=1$, $x_1=\frac{1}{2}$, and two values of the mass ratio: $m_1/m_2=0.5$ and 2. Here, $D_{ij}(1)$ and $D_1^T(1)$ refer to the values of $D_{ij}$ and $D_1^T$ for elastic collisions ($\al=1$). The expressions of $D_1^T$, $D_{11}$ and $D_{12}$ are provided by Eqs.\ \eqref{6.2}--\eqref{6.5}. We observe first that the deviation of the diffusion $D_{ij}$ and thermal diffusion $D_1^T$ coefficients with respect to their forms for elastic collisions (molecular mixtures) is in general significant, as expected. The departure from unity appears even for relatively moderate dissipation (let's say, $\al \simeq 0.8$). Figure~\ref{diffusion} also shows that the coefficients $D_{ij}$ and $D_1^T$ (scaled with their elastic values) exhibit a monotonic dependence on inelasticity, regardless the value of the mass ratio: they increase with increasing dissipation (or equivalently, decreasing $\al$). Thus, inelasticity enhances the mass transport of species. This monotonic behavior found for dilute mixtures is not kept at finite densities ($\phi\neq 0$) since, depending on the value of the mass ratio, the scaled coefficient $D_{11}(\al)/D_{11}(1)$ may exhibit a non-monotonic dependence on $\al$ (see Fig.\ 5.5 of Ref.\ \cite{G19}). An important target of Fig.~\ref{diffusion} is to illustrate the impact of energy non-equipartition on the transport coefficients. The dashed (for $m_1/m_2=0.5$) and dash-dotted (for $m_1/m_2=4$) lines refer to the results obtained for the three coefficients by assuming the equality of the partial temperatures ($T_1^{(0)}=T_2^{(0)}$). The expressions of these coefficients in this approximation are given by Eqs.\ \eqref{6.10}--\eqref{6.11}. As said in the Introduction section of this review, most of the previous studies reported in the granular literature on mixtures were based on this equipartition assumption \cite{JM89,AW98,WA99,Z95,SGNT06,SNTG09}. Figure~\ref{diffusion} highlights the significant effect of energy nonequipartition on mass transport, specially for strong inelasticity. The impact of different partial temperatures ($T_1^{(0)}\neq T_2^{(0)}$) on diffusion coefficients is not only quantitative but also in some cases qualitative. Thus, for instance, while $D_{11}(\al)>D_{11}(1)$ for $m_1/m_2=4$ when energy nonequipartition is accounted for, the opposite [$D_{11}(\al)<D_{11}(1)$] occurs when energy equipartition is assumed. A similar behavior exhibits the coefficient $D_{12}$ in the case $m_1/m_2=0.5$. As expected, the important differences found between both theories (with and without energy equipartition) clearly shows that the effect of different species' granular temperatures cannot be neglected in the study of transport properties in granular mixtures. This conclusion contrasts with the results derived by Yoon and Jenkins \cite{YJ06} who conclude that segregation is not greatly affected by the difference in temperatures of the two species, at least when the particles of both species are nearly elastic and their masses or sizes do not differ by too much. On the other hand, other studies \cite{TAH03,BRM05,ATH06,BRM06,G06,G08a,G09,G11} have shown the important influence of energy nonequipartition on segregation.

As a complement of Fig.\ ~\ref{diffusion}, we consider now the tracer limit $x_1\to 0$. In this limit case, $D_{22}\propto x_1$ and $D_1^T\propto x_1$ and so, both coefficients vanish when one of the species of the mixture is present in tracer concentration. The expression of the tracer diffusion coefficient $D_{11}$ simply reads
\beq
\label{6.11.1}
D_{11}=\frac{\gamma_1}{\nu_D^*-\frac{1}{2}\zeta_0^*},
\eeq
where in the tracer limit
\beq
\label{6.11.2}
\nu_D^*=\frac{\sqrt{2}\pi^{(d-1)/2}}{d\Gamma\left(\frac{d}{2}\right)}
\sqrt{\mu_{21}\left(1+\frac{m_2}{m_1}\gamma_1\right)}, \quad \zeta_0^*=\frac{\pi^{(d-1)/2}}{d\Gamma\left(\frac{d}{2}\right)}\left(\frac{\sigma_2}
{{\sigma}_{12}}\right)^{d-1}\mu_{21}^{-1/2}
\left(1-\al_{22}^2\right).
\eeq
Figure \ref{tracer} shows the $\al$-dependence of the (scaled) tracer diffusion coefficient $D_{11}(\al)/D_{11}(1)$ for $d=3$, $\sigma_1/\sigma_2=2$, and $m_1/m_2=8$. As in Fig.\ ~\ref{diffusion}, the influence of energy nonequipartition on $D_{11}$ is quite relevant, specially at strong inelasticity. Moreover, the comparison with the simulation results obtained from the DSCM method shows an excellent agreement, showing again the accuracy of the first Sonine approximation to $D_{11}$.

To end this section, the shear viscosity coefficient $\eta$ is considered. Figure \ref{viscosity} shows the shear viscosity coefficient (scaled with respect to its value for elastic collisions) for a three-dimensional moderately dense binary mixture ($\phi=0.1$) with the same parameters as in Fig.~\ref{diffusion}. Although the qualitative behavior of $\eta(\al)/\eta(1)$ is quite similar with and without energy equipartition (there is a monotonic decrease in shear viscosity as inelasticity increases in all the cases), there are important quantitative discrepancies between both theories specially in the case $m_1/m_2=4$. To complement Fig.\ \ref{viscosity}, we plot in Fig.\ \ref{viscosity2d} $\eta(\al)/\eta(1)$ for a two-dimensional ($d=2$) dilute ($\phi=0$) binary mixture with $x_1=\frac{1}{2}$ and $m_1/m_2=(\sigma_1/\sigma_2)^2$ (particles of the same mass density). We observe good agreement with Monte Carlo simulations when energy nonequipartiton is accounted for in the theory.
Thus, as in the case of the diffusion coefficients and based on the findings of Figs.\ \ref{viscosity} and \ref{viscosity2d}, we can conclude that a reliable kinetic theory for granular mixtures needs to take into account nonequipartition effects in momentum transport.

\section{First-order contributions to the partial temperatures. Influence on the bulk viscosity}
\label{sec7}

As mentioned in section \ref{sec1}, the presence of a divergence $\nabla \cdot \mathbf{U}$ of the flow velocity  in a mixture induces nonzero first-order contributions $T_i^{(1)}$ to the partial temperatures. This breakdown of the energy equipartition is additional to the one appearing in the HCS which is only due to the inelastic character of the binary collisions. In fact, $T_i^{(1)}\neq 0$ even in the case of molecular \emph{dense} mixtures, namely, a dense hard-sphere mixture with elastic collisions \cite{LCK83,KS79a,KS79b}.

The fact that the partial temperatures $T_i^{(1)}$ are proportional to $\nabla \cdot \mathbf{U}$ gives rise to a contribution to the bulk viscosity $\eta_\text{b}$ coming from these temperatures. In addition, for granular mixtures, the temperatures $T_i^{(1)}$ are also involved in the evaluation of the first-order contribution $\zeta_U$ (proportionality coefficient between $\zeta$ and $\nabla \cdot \mathbf{U}$) to the cooling rate. The coupling between $\eta_\text{b}$ and $T_i^{(1)}$ was already recognized by the pioneering works of the Enskog equation for multicomponent molecular gases \cite{LCK83,KS79a,KS79b}.

According to the definition \eqref{2.8} of $T_i$, its first-order contribution is
\beq
\label{7.1}
T_i^{(1)}=\frac{m_i}{d n_i}\int d\mathbf{v}\; V^2 f_i^{(1)}(\mathbf{V}),
\eeq
where $f_i^{(1)}(\mathbf{r}, \mathbf{V}; t)$ is given by Eq.\ \eqref{5.8}. Since $T_i^{(1)}$ is a scalar, it can be only coupled to $\nabla \cdot \mathbf{U}$ because $\nabla n$ and $\nabla T$ are vectors and the tensor $\partial_\lambda U_\beta+\partial_\beta U_\lambda-(2/d)\delta_{\lambda\beta}\nabla \cdot \mathbf{U}$ is a traceless tensor. As a consequence, $T_i^{(1)}$ can be written as
\beq
\label{7.2}
T_i^{(1)}=\varpi_i \nabla \cdot \mathbf{U}, \quad \varpi_i=\frac{m_i}{d n_i}\int d\mathbf{v}\; V^2 \mathcal{D}_i(\mathbf{V}),
\eeq
where the scalar quantities $\mathcal{D}_i(\mathbf{V})$ obey the following set of coupled linear integral equations \cite{GDH07}:
\beq
\label{7.3}
\frac{1}{2}\zeta^{(0)}\frac{\partial}{\partial {\bf V}}\cdot \left({\bf V}\mathcal{D}_{i}\right)+\frac{1}{2}\zeta^{(0)}\mathcal{D}_{i}
+\frac{1}{2}\zeta^{(1,1)}\frac{\partial}{\partial {\bf V}}\cdot \left({\bf V}f_{i}^{(0)}\right)-\sum_{j=1}^s\chi_{ij}\left(J_{ij}^{\text{B}}[\mathcal{D}_{i},f_j^{(0)}]+
J_{ij}^{\text{B}}[f_i^{(0)},\mathcal{D}_{j}]\right)=D_{i}.
\eeq
Here, $\zeta^{(0)}=\zeta_1^{(0)}=\zeta_2^{(0)}$ is obtained from Eq.\
\eqref{3.7} by replacing $f_i$ and $f_j$ by $f_i^{(0)}$ and $f_j^{(0)}$, respectively. Moreover, $J_{ij}^{\text{B}}$ is the Boltzmann collision operator \eqref{3.2}, the coefficient $\zeta^{(1,1)}$ is given in terms of $\mathcal{D}_i$ as \footnote{In this section, it is understood that $\chi_{ij}$ is evaluated at the zeroth-order approximation}
\beq
\label{7.4}
\zeta^{(1,1)}=\frac{1}{nT}\frac{\pi^{(d-1)/2}}{d\Gamma\left(\frac{d+3}{2}\right)}\sum_{i=1}^s\sum_{j=1}^s \sigma_{ij}^{d-1} \chi_{ij} m_{ij} (1-\al_{ij}^2)\int d\mathbf{v}_1\int d\mathbf{v}_2 \; g_{12}^3\; f_i^{(0)}(\mathbf{V}_1)\mathcal{D}_j (\mathbf{V}_2),
\eeq
and the homogeneous term $D_{i}(\mathbf{V})$ is \cite{GDH07}
\beq
\label{7.5}
D_i\left(\mathbf{V}\right)=\frac{1}{2}\Bigg[\frac{2}{d}\left(1-p^*\right)-\zeta^{(1,0)}\Bigg]
\frac{\partial}{\partial \mathbf{V}}\cdot \left(\mathbf{V}f_i^{(0)}\right)-f_i^{(0)}+\sum_{j=1}^s \Bigg(n_j \frac{\partial f_i^{(0)}}
{\partial n_j} +
\frac{1}{d}\mathcal{K}_{ij,\beta}\left[\frac{\partial f_i^{(0)}}{\partial V_\beta}\right]\Bigg).
\eeq
In Eq.\ \eqref{7.5}, $p^*\equiv p/(n T)$ is the (reduced) hydrostatic pressure [$p$ is given by Eq.\ \eqref{3.5}],
\beq
\label{7.7}
\zeta^{(1,0)}=-\frac{3\pi^{d/2}}{d^2\Gamma\left(\frac{d}{2}\right)}\sum_{i=1}^s\sum_{j=1}^s x_i n_j \mu_{ji}\sigma_{ij}^d \chi_{ij} \gamma_i(1-\al_{ij}^2),
\eeq
and the collision operator $\boldsymbol{\mathcal{K}}_{ij}[X_j]$ is
\beq
\label{7.8}
\boldsymbol{\mathcal{K}}_{ij}[X_j] =\sigma_{ij}^{d}\chi_{ij} \int \dd \mathbf{v}_{2}\int \dd\widehat{\boldsymbol {\sigma
}}\Theta (\widehat{\boldsymbol {\sigma}} \cdot
\mathbf{g}_{12})(\widehat{\boldsymbol {\sigma }}\cdot
\mathbf{g}_{12})\widehat{\boldsymbol {\sigma}}\left[ \alpha_{ij}
^{-2}f_i^{(0)}(\mathbf{v}_{1}'')X_j(\mathbf{v}_{2}'')+f_i^{(0)}(\mathbf{v}_{1})X_j(\mathbf{v}_{2})\right].
\eeq

As said before, as a byproduct the calculation of $T_i^{(1)}$ allows us to compute the first-order contribution to the cooling rate
\beq
\label{7.9}
\zeta=\zeta^{(0)}+\zeta_U \nabla \cdot \mathbf{U}, \quad \zeta_U=\zeta^{(1,0)}+\zeta^{(1,1)},
\eeq
where the coefficients $\zeta^{(1,1)}$ and $\zeta^{(1,0)}$ are defined by Eqs.\ \eqref{7.4} and \eqref{7.7}, respectively.

As in the case of the Navier--Stokes transport coefficients, the evaluation of the first-order contributions $T_i^{(1)}$ requires to solve the integral equations \eqref{7.3}. These equations can be approximately solved by considering the leading Sonine approximation to $\mathcal{D}_i(\mathbf{V})$. Before taking this sort of approximation, it is convenient to prove the solubility condition \eqref{5.7}, or equivalently,
\beq
\label{7.10}
\sum_{i=1}^s\int d\mathbf{v}\; m_i V^2\; {D}_i(\mathbf{V})=0.
\eeq
Upon writing the condition \eqref{7.10} we have taken into account that $\mathcal{D}_i(\mathbf{V})\propto {D}_i(\mathbf{V})$. The constraint \eqref{7.10} yields
\beq
\label{7.11}
\sum_{i=1}^s\; n_i T_i^{(1)}=0,
\eeq
and consequently, the granular temperature $T$ is not affected by the spatial gradients, as expected in the Chapman--Enskog method \cite{CC70}. According to Eq.\ \eqref{7.11}, only $s-1$ partial temperatures $T_i^{(1)}$ are independent. The solubility condition \eqref{7.10} can be verified by using the relation $\sum_i x_i \gamma_i=1$ and the result \cite{GGG19b}
\beqa
\label{7.12}
A_{i}&\equiv& \sum_{j=1}^s\int d\mathbf{v}\; m_i V^2 \mathcal{K}_{ij,\lambda}\left[\frac{\partial f_j^{(0)}}{\partial V_\lambda}\right]\nonumber\\
&=&-
\frac{\pi^{d/2}}{\Gamma\left(\frac{d}{2}\right)}T\sum_{j=1}^s\chi_{ij} n_i n_j \sigma_{ij}^d (1+\al_{ij})\Bigg[3\mu_{ji}(1+\al_{ij})\left(\frac{\gamma_i}{m_i}+\frac{\gamma_j}{m_j}\right)-4\frac{\gamma_i}{m_i}\Bigg].
\eeqa

In the low-density regime ($n_i\sigma_{ij}^d\to 0$), $p^*=1$, $\zeta^{(1,0)}=0$, $\sum_j n_j \partial f_i^{(0)}/\partial n_j-f_i^{(0)}=0$, $\boldsymbol{\mathcal{K}}_{ij}[X_j]=0$, and Eq.\ \eqref{7.5} leads to $D_i(\mathbf{V})=0$. Thus, the homogeneous term $D_i$ vanishes in the integral equation \eqref{7.3} and so, $\mathcal{D}_i=0$. This implies that the first-order contributions $\varpi_i$ to the partial temperatures vanish for \emph{dilute} granular mixtures \cite{GD02,GMD06,SGNT06,GM07,SNTG09}.

\subsection{Bulk viscosity coefficient}

The bulk viscosity $\eta_\text{b}$ is defined through the constitutive equation \eqref{5.10}. This transport coefficient plays a relevant role in problems where the gas density varies in the flow motion; it represents an additional resistance to contraction or expansion. Since $\eta_\text{b}$ has only collisional contributions, its form can be identified by expanding the collisional transfer contribution \eqref{2.19} to the pressure tensor to first order in the spatial gradients. The expression of $\eta_\text{b}$ can be written as \cite{GGG19b}
\beq
\label{7.13}
\eta_\text{b}=\eta_\text{b}'+\eta_\text{b}'',
\eeq
where
\beq
\label{7.14}
\eta_\text{b}^{\prime}=\frac{\pi^{(d-1)/2}}{\Gamma\left(\frac{d+3}{2}\right)}\frac{d+1}{2d^{2}}\sum_{i=1}^{s}\sum_{j=1}^{s}m_{ij}
\left(1+\alpha_{ij}\right)\chi_{ij}\sigma_{ij}^{d+1}\int d\mathbf{v}_1\int d\mathbf{v}_{2}f_{i}^{(0)}(\mathbf{V}_{1})f_{j}^{(0)}(\mathbf{V}_{2})g_{12},
\eeq
and
\beq
\label{7.15}
\eta_\text{b}''=-\frac{\pi^{d/2}}{d\Gamma\left(\frac{d}{2}\right)}\sum_{i=1}^{s}\sum_{j=1}^{s}\mu_{ji}
\left(1+\alpha_{ij}\right)\chi_{ij} n_i n_j\sigma_{ij}^{d}\varpi_i.
\eeq
While the first contribution $\eta_\text{b}^{\prime}$ to the bulk viscosity is given in terms of the zeroth-order distributions $f_i^{(0)}$, the second contribution $\eta_\text{b}''$ is given in terms of the first-order contributions $\varpi_i$ to the partial temperatures. Although this second contribution has been in fact neglected in several previous works \cite{G19,GDH07,GHD07} on dense granular mixtures, as said before it was already computed in the pioneering studies on molecular hard-spheres mixtures \cite{KS79a,KS79b}. The impact of $\eta_\text{b}''$ on $\eta_b$ will be assessed in the next subsection when we estimate $\varpi_i$ by taking the corresponding leading Sonine approximation to $\mathcal{D}_i$. Note that the expression \eqref{7.13} for the bulk viscosity can be written as
\beq
\label{7.15.1}
\eta_\text{b}=\sum_{i=1}^s\; \eta_\text{b}^{i},
\eeq
where the forms of the partial shear viscosity coefficients $\eta_\text{b}^{i}$ can be easily obtained from Eqs.\ \eqref{7.14} and \eqref{7.15}. These forms could provide some insight into a shear-induced segregation problem.

An accurate estimate of the first contribution $\eta_\text{b}'$ to the bulk viscosity is obtained by replacing $f_i^{(0)}(\mathbf{V})$ by the Maxwellian distribution \eqref{6.1}. With this approximation, $\eta_\text{b}'$ is  \cite{GDH07}
\beq
\label{7.17}
\eta_\text{b}'=\frac{\pi^{(d-1)/2}}{d^2\Gamma\left(\frac{d}{2}\right)}
v_\text{th}\sum_{i=1}^{2}\sum_{j=1}^{2}m_{ij}
\left(1+\alpha_{ij}\right)\chi_{ij} n_i n_j \sigma_{ij}^{d+1}\left(\frac{\theta_i+\theta_j}
{\theta_i\theta_j}\right)^{1/2},
\eeq
where $\theta_i$ is given in Eq.\ \eqref{3.15}.

\subsection{Leading Sonine approximation to $\varpi_i$}

The coefficient $\varpi_i$ is defined by Eq.\ \eqref{7.2}. To estimate it, we take the following Sonine approximation to $\mathcal{D}_i(\mathbf{V})$:
\beq
\label{7.18}
\mathcal{D}_i(\mathbf{V})\rightarrow f_{i,\text{M}}(\mathbf{V})W_i(\mathbf{V})\frac{\varpi_i}{T_i^{(0)}}, \quad
W_i(\mathbf{V})=\frac{m_iV^2}{2T_i^{(0)}}-\frac{d}{2}.
\eeq
The coefficients $\varpi_i$ can be determined by substituting \eqref{7.18} into the integral equations \eqref{7.3}, multiplying them with the polynomial $W_i(\mathbf{V})$, and integrating over the velocity. The procedure is large but straightforward. Technical details for multicomponent mixtures can be found in Ref.\ \cite{GGG19b}. Here, we focus on the case of a binary mixture ($s=2$). In this case, $\varpi_2=-(x_1/x_2)\varpi_1$ and   $\varpi_1=(T/(n\sigma_{12}^{d-1}v_\text{th}))\varpi_1^*$, where
\beq
\label{7.19}
\varpi_1^*=\frac{\frac{2}{d}\gamma_1\left(1-p^*\right)-\gamma_1\zeta^{(1,0)}-\phi \frac{\partial \gamma_1}{\partial \phi}-\frac{A_1}{d^2 n_1 T}}{\omega_{11}^*-\frac{x_1}{x_2}\omega_{12}^*+\frac{1}{2}\zeta_0^*+\gamma_1\left(\xi_1^*-\frac{x_1}{x_2}\xi_2^*\right)}.
\eeq
In Eq.\ \eqref{7.19}, we have introduced the dimensionless quantities
\beq
\label{7.20}
\xi_i^*=\frac{3\pi^{(d-1)/2}}{d\Gamma\left(\frac{d}{2}\right)\gamma_i}\sum_{j=1}^2 x_i x_j \left(\frac{\sigma_{ij}}{\sigma_{12}}\right)^{d-1}
\chi_{ij} \frac{m_{ij}}{\overline{m}}(1-\al_{ij}^2) \left(\theta_i+\theta_j\right)^{1/2}\theta_i^{-3/2}\theta_j^{-1/2}, \quad (i=1,2),
\eeq
\beqa
\label{7.21}
\omega_{11}^*&=&-\frac{\pi^{(d-1)/2}}{d\Gamma\left(\frac{d}{2}\right)\gamma_1}
\Bigg\{3\sqrt{2}
x_1\left(\frac{\sigma_{1}}{\sigma_{12}}\right)^{d-1} \mu_{12}\chi_{11} \theta_1^{-3/2}
\left(1-\alpha_{11}^2\right)-2x_2 \mu_{12}\mu_{21} \chi_{12} \left(1+\alpha_{12}\right)\nonumber\\
& & \times \left(\theta_1+\theta_2\right)^{-1/2}
\theta_1^{-3/2} \theta_2^{-1/2}
\Big[3\mu_{21}\left(1+\alpha_{12}\right)\left(\theta_1+\theta_2\right)-
2\left(2\theta_1+3\theta_2\right)\Big]\Bigg\},
\eeqa
\beqa
\label{7.22}
\omega_{12}^*&=&\frac{2\pi^{(d-1)/2}}{d \Gamma\left(\frac{d}{2}\right)\gamma_2} x_2 \mu_{12}\mu_{21} \chi_{12}
\left(1+\alpha_{12}\right)\left(\theta_1+\theta_2\right)^{-1/2}
\theta_1^{-1/2}\theta_2^{-3/2}\nonumber\\
& & \times \Big[3\mu_{21}\left(1+\alpha_{12}\right)\left(\theta_1+\theta_2\right)-2\theta_2\Big],
\eeqa
where we recall that $\overline{m}=(m_1+m_2)/2$ for a binary mixture.

Equation \eqref{7.19} clearly shows that the coefficient $\varpi_1^*$ displays a quite nonlinear dependence on the parameter space of the mixture. In the low-density regime ($\phi=0$), $p^*=1$, $\zeta^{(1,0)}=0$, and $A_1=0$ so that $B_1=0$ and $\varpi_1^*=0$. This is the expected result for \emph{dilute} granular mixtures \cite{GD02,GMD06,GM07}. However, $\varpi_1^* \neq 0$ in binary granular suspensions at low-density \cite{GKG20,GGG20} and confined quasi-two-dimensional dilute granular mixtures \cite{GBS21}.

\begin{figure}
\includegraphics[width=0.4\textwidth]{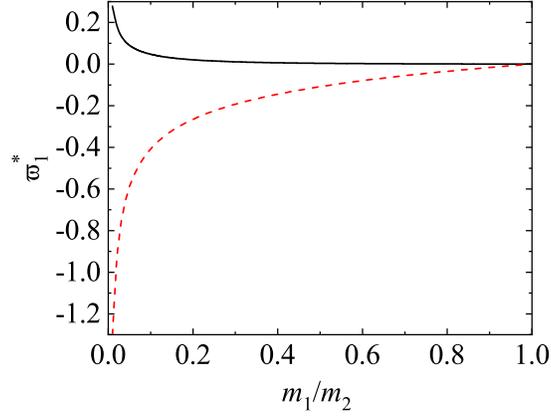}
\caption{Dependence of the (reduced) coefficient $\varpi_1^*$ versus the mass ratio $m_1/m_2$ for a molecular binary mixture of hard disks ($d=2$) when $x_1=\frac{1}{2}$, $\phi=0.25$ and $\sigma_1/\sigma_2=(m_1/m_2)^{1/2}$. The solid and dashed lines correspond to the results obtained here and those reported by Jenkins and Mancini \cite{JM87}, respectively.
\label{jenkins}}
\end{figure}
\begin{figure}
\includegraphics[width=0.4\textwidth]{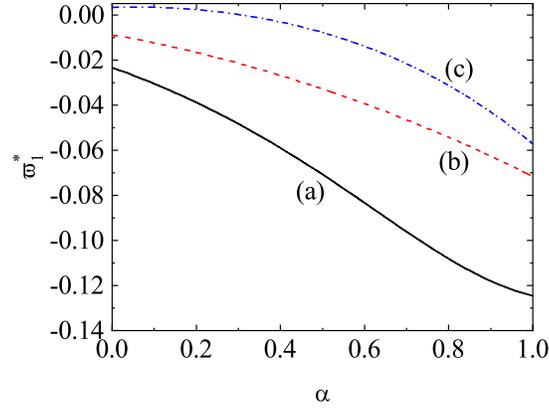}
\caption{Dependence of the (reduced) coefficient $\varpi_1^*$ versus the (common) coefficient of restitution $\al$ for hard spheres ($d=3$) with $x_1=\frac{1}{2}$, $\phi=0.25$, and $\sigma_1/\sigma_2=(m_1/m_2)^{1/3}$. Three different values of the mass ratio are considered: $m_1/m_2=0.5$ (a), $m_1/m_2=2$ (b), and $m_1/m_2=5$ (c).
\label{om1}}
\end{figure}

Another simple but interesting case corresponds to molecular mixtures of dense hard-spheres. In this case ($\al_{11}=\al_{22}=\al_{12}=1$), $\zeta_0^*=\zeta^{(1,0)}=\xi_i^*=0$, $\gamma_i=1$, $\theta_1=2\mu_{12}$, $\theta_2=2\mu_{21}$, and $\varpi_1^*$ becomes
\beq
\label{7.23}
\varpi_1^*=\frac{4\pi^{d/2}}{d^2\Gamma\left(\frac{d}{2}\right)}\frac{n_2\sigma_{12}^d\chi_{12}\big(x_2\mu_{21}-x_1\mu_{12}\big)
+\frac{1}{2}x_2\left(n_1\sigma_1^d \chi_{11}-n_2\sigma_2^d \chi_{22}\right)}{\omega_{11,\text{el}}^{*}-\frac{x_1}{x_2}\omega_{12,\text{el}}^{*}},
\eeq
where the expressions of $\omega_{11,\text{el}}^{*}$ and $\omega_{12,\text{el}}^{*}$ are easily obtained from Eqs.\ \eqref{7.21} and \eqref{7.22}, respectively, by considering elastic collisions. The expression \eqref{7.23} agrees with the one obtained many years ago by Karkheck and Stell \cite{KS79b} for a hard-sphere binary mixture ($d=3$). On the other hand, for a two-dimensional system ($d=2$), Eq.\ \eqref{7.23} differs from the one derived by Jenkins and Mancini \cite{JM87} for nearly elastic hard disks. As recognized by the authors of this paper, given that their prediction on $\varpi_1^*$ was derived by assuming Maxwellian distributions for each species, a more accurate expression of $\varpi_1^*$ is obtained when one evaluates this coefficient from the first-order distribution of the Chapman--Enskog solution. In particular, Eq.\ \eqref{7.23} takes into account not only the different centers $\mathbf{r}$ and $\mathbf{r}\pm \boldsymbol{\sigma}$ of the colliding spheres in the Enskog collision operator (this is in fact the only ingredient accounted for in Ref.\ \cite{JM87} for getting $\varpi_1^*$) but also the form of the first-order distribution functions $f_i^{(1)}$ given by Eq.\ \eqref{5.8}. Moreover, while $\varpi_1^*\to 0$ for vanishing densities ($\phi\to 0$), the results found by Jenkins and Mancini \cite{JM87} predict a nonvanishing $\varpi_1^*$ for dilute binary mixtures if $m_1\neq m_2$. This result contrasts with those obtained for molecular mixtures \cite{KS79a,KS79b}.

To illustrate the differences between the results obtained in Ref.\ \cite{JM87} and those derived here for disks, Fig.~\ref{jenkins} shows $\varpi_1^*$ versus $m_1/m_2$ when $x_1=\frac{1}{2}$, $\phi=0.25$ and
$\sigma_1/\sigma_2=(m_1/m_2)^{1/2}$ (i.e., when the disks are made of the same material). In the case of disks \cite{JM87},
\beq
\label{7.25}
\chi_{ij}=\frac{1}{1-\phi}+\frac{9}{16}\frac{\phi}{(1-\phi)^2}\frac{\sigma_i \sigma_j M_1}{\sigma_{ij}M_2},
\eeq
where $\phi=\sum_i n_i \pi \sigma_i^2/4$ is the solid volume fraction for disks and we recall that $M_\ell=\sum_i x_i \sigma_i^\ell$. It is quite apparent the differences found between both theories, specially for disparate masses.

\begin{figure}
\includegraphics[width=0.4\textwidth]{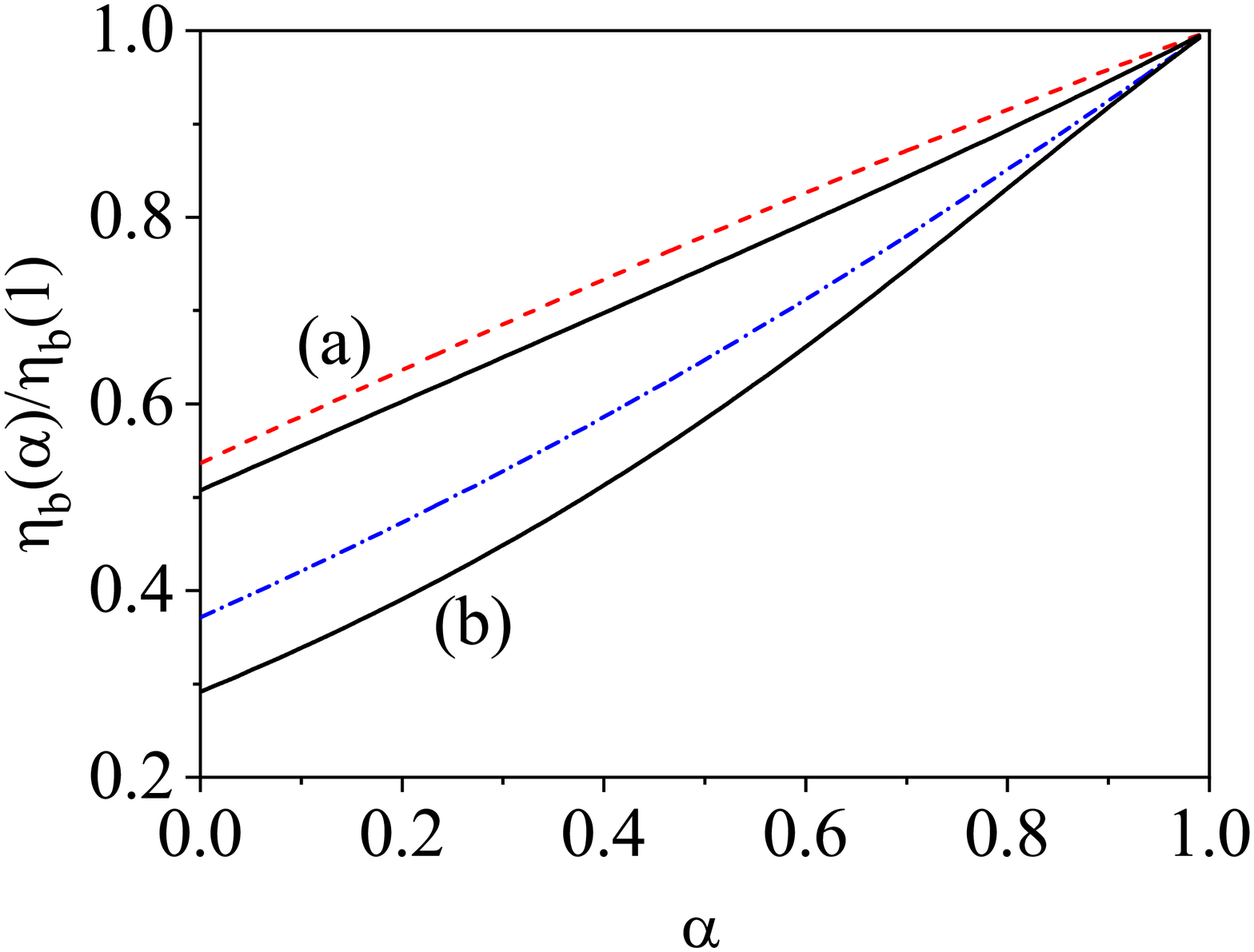}
\caption{Dependence of the (reduced) bulk viscosity $\eta_\text{b}(\al)/\eta_\text{b}(1)$ on the (common) coefficient of restitution $\al$ for a granular binary mixture of hard spheres ($d=3$) with $x_1=\frac{1}{2}$, $\phi=0.25$, $\sigma_1/\sigma_2=2$, and two different values of the mass ratio: $m_1/m_2=0.5$ (a) and $m_1/m_2=5$ (b). The solid lines are the results obtained here while the dashed lines correspond to the results obtained for the (reduced) bulk viscosity when the contribution $\eta_\text{b}''$ to $\eta_\text{b}$ is neglected.
\label{etab}}
\end{figure}
\begin{figure}
\includegraphics[width=0.4\textwidth]{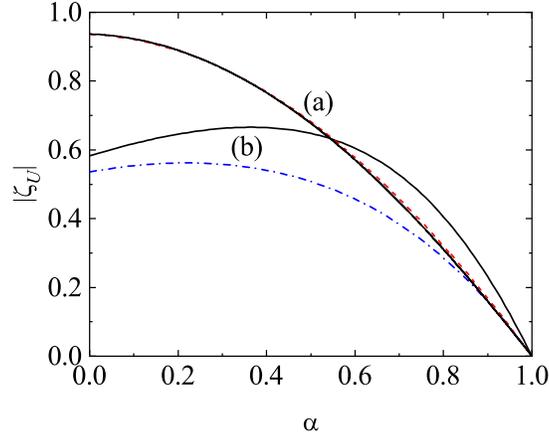}
\caption{Dependence of the magnitude of the (reduced) first-order contribution $\zeta_U$ to the cooling rate
on the (common) coefficient of restitution $\al$ for a granular binary mixture of hard spheres ($d=3$) with $x_1=\frac{1}{2}$, $\phi=0.25$, and $\sigma_1/\sigma_2=2$, and two different values of the mass ratio: $m_1/m_2=0.5$ (a) and $m_1/m_2=5$ (b). The solid lines are the results obtained here while the dashed lines correspond to the results obtained for the (reduced) cooling rate when the contribution $\zeta^{(1,1)}$ to $\zeta_U$ is neglected.
\label{zetau}}
\end{figure}

For inelastic collisions, Fig.~\ref{om1} illustrates the dependence of $\varpi_1^*$ on the (common) coefficient of restitution $\al$ for a binary mixture of hard spheres ($d=3$) with $x_1=\frac{1}{2}$, $\phi=0.25$, $m_1/m_2=(\sigma_1/\sigma_2)^3$ and three different values of the mass ratio: $m_1/m_2=0.5$, 2 and 5. We observe first that $\varpi_1^*$ is significantly affected by inelasticity, specially for high mass ratios. With respect to the effect of the mass ratio on $\varpi_1^*$, we see that this coefficient decreases (increases) with increasing inelasticity when $m_1/m_2>1$ ($m_1/m_2<1$). As expected, Fig.~\ref{om1} also shows that the magnitude of $\varpi_1^*$ is in general quite small in comparison with the remaining transport coefficients.

\subsection{Influence of $T_i^{(1)}$ on the bulk viscosity and the cooling rate}

According to Eqs.\ \eqref{7.13}--\eqref{7.15}, the coefficient $\varpi_1^*$ is involved in the contribution $\eta_\text{b}''$ to the bulk viscosity $\eta_\text{b}$. To assess the impact of the first-order contributions to the partial temperatures on the bulk viscosity, we plot in Fig.~\ref{etab} the (reduced) bulk viscosity $\eta_\text{b}(\al)/\eta_\text{b}(1)$ as a function of the (common) coefficient of restitution $\al$. As in Fig.~\ref{om1}, $x_1=\frac{1}{2}$, $\phi=0.25$ and $m_1/m_2=(\sigma_1/\sigma_2)^3$. Two different mass ratios are studied: $m_1/m_2=0.5$ and 5. The value of the (reduced) bulk viscosity when the coefficient $\varpi_1^*$ is neglected (dashed lines) is also plotted for the sake of comparison. Although both results (with and without the contribution coming from $\eta_\text{b}''$) agree qualitatively, Fig.~\ref{etab} highlights that the impact of $\varpi_1^*$ on the bulk viscosity cannot be neglected for high mass ratios and strong dissipation (let's say, for instance, $\al\lesssim 0.5$).

Finally, Fig.~\ref{zetau} shows the $\alpha$-dependence of the first-order contribution $\zeta_U$ to the cooling rate. This coefficient is defined by Eq.\ \eqref{7.9} where $\zeta^{(1,1)}$ is
\beq
\label{7.26}
\zeta^{(1,1)}=\Big(\xi_1^*-\frac{x_1}{x_2}\xi_2^*\Big)\varpi_1^*.
\eeq
The coefficients $\xi_1^*$ and $\xi_2^*$ are given by Eq.\ \eqref{7.20}. As Fig.~\ref{etab}, Fig.~\ref{zetau} highlights that the influence of $\varpi_1^*$ turns out to be relevant for strong inelasticities and high mass ratios.

\section{Summary and concluding remarks}
\label{sec8}

The primary objective of this review has been to analyze the influence of energy nonequipartition on the transport coefficients of an $s$-component granular mixture. Granular mixtures have been modeled here as a collection of inelastic hard spheres of masses $m_i$ and diameters $\sigma_i$ ($i=1,\cdots,s$). We have also assumed that spheres are completely smooth so that the inelasticity of collisions is only accounted for by the (positive) constant coefficients of normal restitution $\al_{ij}\leq 1$. At a kinetic level, all the relevant information on the state of the mixture is given through the knowledge of the one-particle velocity distribution functions $f_i(\mathbf{r},\mathbf{v};t)$ of each species. At moderate densities, the distributions $f_i$ verify the set of $s$-coupled Enskog kinetic equations.

The study of the influence of different partial temperatures on transport has been carried out in two different steps. First, we have widely analyzed the failure of energy equipartition in granular mixtures in the HCS, namely, a homogeneous \emph{freely} cooling state. The understanding of this simple situation is crucial because the HCS plays the role of the reference state in the Chapman--Enskog solution to the Enskog equation. Assuming the scaling solution \eqref{3.10} for the distributions $f_i$, the temperature ratios $\gamma_i\equiv T_i^{(0)}/T$ have been determined from the consistency conditions \eqref{3.11} for the HCS: $\zeta_1=\zeta_2=\ldots=\zeta$. To estimate the partial cooling rates $\zeta_i$, the leading Sonine approximation \eqref{3.14} to the scaled distributions $\varphi_i$ have been considered. This approximation also  involves the calculation of the second Sonine coefficients $a_2^{(i)}$.

The temperature ratios $\gamma_i$ and the Sonine coefficients $a_2^{(i)}$ have been both approximately determined by neglecting nonlinear terms in $a_2^{(i)}$ in the corresponding collisional integrals. These theoretical predictions have been tested via a comparison with Monte Carlo (DSMC) and MD simulations for conditions of practical interest. Comparison between DSMC simulations and theory shows in general an excellent agreement; more discrepancies are observed in the case of MD simulations, specially for high volume fractions and/or strong dissipation. This disagreement is a clear indication of the limitations of the Enskog theory in these ranges of values of volume fraction (or density) and/or inelasticity. On the other hand, the good agreement found with the DSMC results reinforces the reliability and accuracy of the approximate analytical predictions even for disparate mass and diameter ratios and/or small values of the coefficients of restitution. As expected, the deviations from the energy equipartition ($T_i^{(0)}/T\neq 1$) can be weak or strong depending on the mechanical differences between the different species of the mixture and the degree of inelasticity in collisions.

Once the dependence of the ratios $T_i^{(0)}/T$ on the parameter space of the mixture has been characterized, the next step has been to study the influence of nonequipartition on the Navier--Stokes transport coefficients. This study is relevant since many of previous attempts reported in the literature \cite{JM89,Z95,AW98,WA99} for obtaining the transport coefficients of granular mixtures had assumed the equality of the partial temperatures ($T_1^{(0)}=T_2^{(0)}=\ldots=T_s^{(0)}=T$). Thus, in contrast with the conclusions reached in previous works \cite{YJ06}, our analysis shows that the impact of different partial temperatures on transport is in general quite significant, as has been clearly illustrated in Figs.~\ref{diffusion}--\ref{viscosity2d} for the diffusion and shear viscosity coefficients.

As a second step in the paper, we have also analyzed the failure of energy equipartition due to the presence of spatial gradients in the system. More specifically, in a dense mixture a nonzero divergence $\nabla \cdot \mathbf{U}$ of the flow velocity field induces a nonzero contribution to the partial temperatures $T_i$. This first-order contribution $T_i^{(1)}$ to $T_i$ is not generic of dense granular mixtures since it is also present for elastic collisions. In fact, previous pioneering works \cite{LCK83,KS79a,KS79b} on dense hard-sphere molecular mixtures ($\al_{ij}=1$) determine these coefficients in terms of the parameters of the mixture. Here, we have extended those calculations to the case of granular mixtures ($\al_{ij}<1$).

A careful analysis of the first-order Chapman--Enskog solution to the Enskog equation shows that the coefficients $T_i^{(i)}$ are involved in the evaluation of the bulk viscosity $\eta_\text{b}$ (proportionality coefficient between the collisional part $\mathsf{P}^c$ of the pressure tensor and $\nabla \cdot \mathbf{U}$) and the first-order contribution $\zeta_U$ to the cooling rate $\zeta$ (proportionality coefficient between $\zeta$ and $\nabla \cdot \mathbf{U}$). Thus, although the coefficients $T_i^{(1)}$ are not hydrodynamic quantities (in a similar way to the partial temperatures $T_i^{(0)}$), they contribute to the value of the bulk viscosity. On the other hand, their impact on transport is in general smaller than the one found in the case of the zeroth-order contributions $T_i^{(0)}$ to the partial temperatures. Our results indicate that the effect of $T_i^{(1)}$ on both $\eta_\text{b}$ and $\zeta_U$ is only relevant for high mass ratios and strong dissipation [see Figs.~\ref{etab} and \ref{zetau}]. In this context, we can conclude that previous expressions of the bulk viscosity and cooling rate for dense granular mixtures \cite{GDH07,GHD07,MGH12} (which implicitly neglect the first-order contributions $T_i^{(1)}$) must incorporate the contributions coming from $T_i^{(1)}$ when the masses of the species are disparate and/or the degree of collisional dissipation turns out to be important.

Granular hydrodynamics derived from hard-sphere models have been shown to be useful in the description of numerous industrial processes involving solid particles. Of particular relevance are high-speed, gas-solid flows, and fluidized beds. Such descriptions are now standard features of commercial and research codes. Since those codes rely upon accurate expressions of the Navier--Stokes transport coefficients, it is quite apparent that a first-order objective is to guarantee a reliable theoretical treatment. As shown in this paper and previous review works \cite{G08c,G19}, the price of this accurate approach (in contrast to more phenomenological approaches) is an increasing complexity of the expressions derived for the transport coefficients.

As mentioned in section \ref{sec1}, since grains in nature are generally surrounded by a fluid like water or air, a granular mixture is in fact a multiphase system. In this review, the influence of the interstitial fluid on the dynamical properties of the granular mixture has been neglected. A further step is to take into account the presence of the surrounding gas and develop a theory for moderately dense granular suspensions. This will provide a fundamental basis for the application of granular hydrodynamics under realistic conditions. Although some previous attempts based on the introduction of solid-fluid forces \cite{KG13,GKG20,BOB20,OBB20} have been made in the past, it still remains to propose theories where the influence of collisions between the solid particles and molecules of the interestitial fluid is explicitly accounted for in the corresponding kinetic equation. The complexity for considering this type of collisions is  not a real problem for implementation in a code.

As shown along this overview, granular mixtures exhibit a wide range of interesting phenomena for which the Navier--Stokes hydrodynamic equations
can be considered as an accurate and practical tool. However, due to their
complexity, many of their features are not fully understood. Kinetic theory
and hydrodynamics (in the broader sense) can be expected to provide some
insight into the understanding of such complex materials.


\acknowledgments

The authors acknowledge financial support from Grant PID2020-112936GB-I00 funded by MCIN/AEI/ 10.13039/501100011033, and from Grants IB20079 and GR18079 funded by Junta de Extremadura (Spain) and by ERDF A way of making Europe. The research of R.G.G. also has been supported by the predoctoral fellowship BES-2017-079725 from the Spanish Government.










\end{document}